\documentclass[longauth]{aa}  

\usepackage{graphicx}
\usepackage{txfonts}

\usepackage{color}

\newcommand{\etal}{~{et~al.}\ }  

\newcommand{\kms}{km~s$^{-1}$}

\newcommand{\hi}{H{\sc\,i}}
\newcommand{\bm}{beam$^{-1}$}
\newcommand{\dg}{$^{\circ}$}

\usepackage{natbib,twoopt}
\usepackage[breaklinks=true]{hyperref} 
\bibpunct{(}{)}{;}{a}{}{,} 
\makeatletter
\newcommandtwoopt{\citeads}[3][][]{\href{http://adsabs.harvard.edu/abs/#3}%
{\def\hyper@linkstart##1##2{}%
\let\hyper@linkend\@empty\citealp[#1][#2]{#3}}}
\newcommandtwoopt{\citepads}[3][][]{\href{http://adsabs.harvard.edu/abs/#3}%
{\def\hyper@linkstart##1##2{}%
\let\hyper@linkend\@empty\citep[#1][#2]{#3}}}
\newcommandtwoopt{\citetads}[3][][]{\href{http://adsabs.harvard.edu/abs/#3}%
{\def\hyper@linkstart##1##2{}%
\let\hyper@linkend\@empty\citet[#1][#2]{#3}}}
\newcommandtwoopt{\citeyearads}[3][][]%
{\href{http://adsabs.harvard.edu/abs/#3}
{\def\hyper@linkstart##1##2{}%
\let\hyper@linkend\@empty\citeyear[#1][#2]{#3}}}
\makeatother

\usepackage[mathlines,switch]{lineno}

\begin{document}

   \title{First release of Apertif imaging survey data}

   \author{
      Elizabeth~A.~K.~Adams       \inst{\ref{astron} \and \ref{kapteyn}}
\and B.~Adebahr           \inst{\ref{airub}}
\and W.~J.~G.~de~Blok     \inst{\ref{astron} \and \ref{cpt} \and \ref{kapteyn}}
\and H.~Dénes            \inst{\ref{astron}}
\and K.~M.~Hess           \inst{\ref{iaa} \and \ref{astron} \and \ref{kapteyn}}
\and J.~M.~van~der~Hulst  \inst{\ref{kapteyn}}
\and A.~Kutkin            \inst{\ref{astron} \and \ref{lebedev}}
\and D.~M.~Lucero         \inst{\ref{virginiatech}}
\and R.~Morganti   \inst{\ref{astron} \and \ref{kapteyn}}
\and V.~A.~Moss           \inst{\ref{csiro} \and \ref{sydney} \and \ref{astron}}
\and T.~A.~Oosterloo      \inst{\ref{astron} \and \ref{kapteyn}}
\and E.~Orr\'u            \inst{\ref{astron}}
\and R.~Schulz            \inst{\ref{astron}}
\and A.~S.~van~Amesfoort  \inst{\ref{astron}}
\and A.~Berger            \inst{\ref{airub}}
\and O.~M.~Boersma        \inst{\ref{uva}}
\and M.~Bouwhuis          \inst{\ref{nikhef}}
\and R.~van~den~Brink     \inst{\ref{astron} \and \ref{tricas}}
\and W.A.~van~Cappellen   \inst{\ref{astron}}
\and L.~~Connor           \inst{\ref{uva} \and \ref{caltech}}
\and A.~H.~W.~M.~Coolen   \inst{\ref{astron}}
\and S.~Damstra           \inst{\ref{astron}}
\and G.~N.~J.~van~Diepen  \inst{\ref{astron}}
\and T.J.~Dijkema         \inst{\ref{astron}}
\and N.~Ebbendorf         \inst{\ref{astron}}
\and Y.G.~Grange          \inst{\ref{astron}}
\and R.~de~Goei           \inst{\ref{astron}}
\and A.W.~Gunst           \inst{\ref{astron}}
\and H.A.~Holties         \inst{\ref{astron}}
\and B.~Hut               \inst{\ref{astron}}
\and M.V.~Ivashina        \inst{\ref{chalmers}}
\and G.I.G.~J\'ozsa       \inst{\ref{mpifr} \and \ref{rhodes}}
\and J.~van~Leeuwen    \inst{\ref{astron} \and \ref{uva}}
\and G.~M.~Loose      \inst{\ref{astron}}
\and Y.~Maan          \inst{\ref{ncra} \and \ref{astron}}
\and M.~Mancini           \inst{\ref{astron}}
\and A'.~Mika             \inst{\ref{astron}}
\and H.~Mulder            \inst{\ref{astron}}
\and M.~J.~Norden         \inst{\ref{astron}}
\and A.~R.~Offringa       \inst{\ref{astron} \and \ref{kapteyn}}
\and L.~C.~Oostrum        \inst{\ref{astron} \and \ref{uva} \and \ref{escience}}
\and I.~Pastor-Marazuela  \inst{\ref{uva} \and \ref{astron}}
\and D.J.~Pisano          \inst{\ref{wvu} \and \ref{gwac} \and \ref{cpt}}
\and A.A.~Ponomareva      \inst{\ref{oxford}}
\and J.W.~Romein          \inst{\ref{astron}}
\and M~Ruiter             \inst{\ref{astron}}
\and A.P.~Schoenmakers    \inst{\ref{astron}}
\and D.~van~der~Schuur    \inst{\ref{astron}}
\and J.J.~Sluman          \inst{\ref{astron}}
\and R.~Smits             \inst{\ref{astron}}
\and K.J.C~Stuurwold      \inst{\ref{astron}}
\and J.~Verstappen     \inst{\ref{kapteyn} \and \ref{astron}}
\and N.P.E~Vilchez        \inst{\ref{astron}}
\and D.~Vohl            \inst{\ref{uva} \and \ref{astron}}
\and K.J.~Wierenga        \inst{\ref{astron}}
\and S.~J.~Wijnholds      \inst{\ref{astron}}
\and E.E.M.~Woestenburg   \inst{\ref{astron}}
\and A.W.~Zanting         \inst{\ref{astron}}
\and J.~Ziemke            \inst{\ref{astron} \and \ref{oslocit}}
} 

\institute{
ASTRON, the Netherlands Institute for Radio Astronomy, Oude Hoogeveensedijk 4,7991 PD Dwingeloo, The Netherlands\label{astron}
  \and
Kapteyn Astronomical Institute, PO Box 800, 9700 AV Groningen, The Netherlands\label{kapteyn}
  \and
Astronomisches Institut der Ruhr-Universit\"at Bochum (AIRUB), Universit\"atsstrasse 150, 44780 Bochum, Germany\label{airub}
  \and
Dept.\ of Astronomy, Univ.\ of Cape Town, Private Bag X3, Rondebosch 7701, South Africa\label{cpt}
  \and
Instituto de Astrof\'{i}sica de Andaluc\'{i}a (CSIC), Glorieta de la Astronom\'{i}a s/n, 18008 Granada, Spain\label{iaa}
  \and
Astro Space Center of Lebedev Physical Institute, Profsoyuznaya Str. 84/32, 117997 Moscow, Russia\label{lebedev}
  \and
Department of Physics, Virginia Polytechnic Institute and State University, 50 West Campus Drive, Blacksburg, VA 24061, USA\label{virginiatech}
  \and
CSIRO Astronomy and Space Science, Australia Telescope National Facility, PO Box 76, Epping NSW 1710, Australia\label{csiro}
  \and
Sydney Institute for Astronomy, School of Physics, University of Sydney, Sydney, New South Wales 2006, Australia\label{sydney}
  \and
Anton Pannekoek Institute, University of Amsterdam, Postbus 94249, 1090 GE Amsterdam, The Netherlands\label{uva}
  \and
NIKHEF, National Institute for Subatomic Physics, Science Park 105, 1098 XG Amsterdam, Netherlands\label{nikhef}
  \and
Cahill Center for Astronomy, California Institute of Technology, Pasadena, CA, USA\label{caltech}
  \and
Dept.\ of Electrical Engineering, Chalmers University of Technology, Gothenburg, Sweden\label{chalmers}
  \and
Max-Planck-Institut f{\"u}r Radioastronomie, Auf dem H{\"u}gel 69, D-53121 Bonn, Germany\label{mpifr}
  \and
Department of Physics and Electronics, Rhodes University, P.O. Box 94, Makhanda, 6140, South Africa\label{rhodes}
  \and
National Centre for Radio Astrophysics, Tata Institute of Fundamental Research, Pune 411007, Maharashtra, India\label{ncra}
  \and
Netherlands eScience Center, Science Park 402, 1098 XH Amsterdam, The Netherlands \label{escience}
  \and
Department of Physics and Astronomy, West Virginia University, White Hall, P.O. Box 6315, Morgantown, WV 26506, USA\label{wvu}
  \and
Center for Gravitational Waves and Cosmology, West Virginia University, Chestnut Ridge Research Building, Morgantown, WV 26505\label{gwac}
  \and
Oxford Astrophysics, Denys Wilkinson Building, University of Oxford, Keble Rd, Oxford, OX1 3RH, UK\label{oxford}
  \and
Tricas Industrial Design \& Engineering, Zwolle, The Netherlands\label{tricas}
  \and
University of Oslo Center for Information Technology, P.O. Box 1059, 0316 Oslo, Norway\label{oslocit}
}


 
  \abstract
  {
  Apertif is a phased-array feed {system} for the Westerbork Synthesis Radio Telescope, providing forty instantaneous beams { over 300 MHz of bandwidth}.
    A dedicated
  survey program utilizing this upgrade started on 1 July 2019, with the last observations taken on 28 February 2022. The imaging survey component provides 
  radio continuum, polarization,
  and {spectral} line data.
  }
  {
  Public release of data is critical for maximizing the legacy of a survey. 
  Toward that end,
    we describe the release of data products from
  the first year of survey operations, through 30 June 2020. 
   In particular,
  we focus on defining quality control metrics for the processed data products.
  }
  {The Apertif imaging pipeline, {\tt Apercal}, automatically produces non-primary beam corrected continuum images, polarization images and cubes, and uncleaned spectral line and dirty beam cubes
  for each beam of an Apertif imaging observation. For this release, processed data products are
  considered on a beam-by-beam basis within an observation. We validate the continuum images by using metrics that identify deviations
  from Gaussian noise in the residual images. If the continuum image
  passes validation, we release all processed data products for a given beam. We apply
  further validation to the polarization and line data products and provide flags indicating the quality of those data products.}
  {We release all raw observational data from the first year of survey observations, for a total of {221} observations of {160} independent target fields, covering approximately one thousand square degrees of sky. 
  Images and cubes are released on a per beam basis, 
   and 3374 beams (of 7640 considered)  are released. 
  The median noise in the continuum images is 41.4 uJy \bm, with
  a slightly lower median noise of 36.9 uJy \bm\ in the Stokes V polarization image. 
  The median angular resolution is {11.6}\arcsec/$\sin{\delta}$. 
    The median noise for all line cubes, with a spectral resolution of 36.6 kHz, is
   1.6 mJy \bm, corresponding
  to a 3-$\sigma$ \hi\ column density sensitivity of 1.8$\times 10^{20}$ atoms cm$^{-2}$ over 20 \kms\ (for a median angular resolution of 24\arcsec $\times$15\arcsec ). Line cubes at lower frequency  have slightly higher noise values, consistent with the global RFI environment and overall Apertif system performance. We also
  provide primary beam images for each individual Apertif compound beam.
    The data are made accessible using a Virtual Observatory interface and can be queried using a variety of standard tools.
  }
  {}


   \maketitle
%

\section{Introduction}

Large, {untargeted} surveys have always been a driver of new discoveries in astronomy,
across all wavelength bands. Surveying wide areas of sky or large volumes of the Universe provides large samples of astronomical sources, including many rare and interesting objects.

In the radio regime, surveys have generally fallen into two categories in the past,
based on technological limitations.
In the first category,
radio continuum and polarization surveys utilize interferometers
for higher angular resolution
with backends that provide
a large bandwidth but limited spectral resolution. 
The prototypical example of a such survey
that continues to provide a rich legacy dataset is
the NRAO VLA Sky Survey (NVSS) \citepads{1998AJ....115.1693C}.
The subsequent VLA  Faint Images of the Radio Sky at Twenty-Centimeters (FIRST) survey
demonstrates the power of even higher angular resolution \citep{1997ApJ...475..479W}.
More recently the VLA Synoptic Sky Survey
at slightly higher frequencies is pushing to higher angular resolution
and better sensitivity \citepads{2020PASP..132c5001L}, 
{while the Rapid ASKAP Continuum Survey \citepads{2020PASA...37...48M, 2021PASA...38...58H},
{at} slightly lower frequencies, is increasing
the resolution and sensitivity of radio surveys
available in the Southern Hemisphere.}
Complementary surveys at {low} frequencies provide
an important view of the radio continuum sky.
An early example is the Westerbork Northern Sky Survey 
\citepads[WENSS;][]{1997A&AS..124..259R}.
Recently, 
the low-frequency window is expanding with a wealth of new low-frequency radio continuum and polarization surveys, including the LOFAR Two-metre Sky Survey (LoTSS)
\citepads{2019A&A...622A...1S},
 the LOFAR LBA Sky Survey \citep[LoLSS][]{2021A&A...648A.104D},
 {and the GaLactic and Extragalactic All-Sky MWA Survey 
 \citepads[GLEAM][]{2015PASA...32...25W}}.
 
 In the second category,
 spectral line surveys, notably
 of neutral hydrogen (\hi) at 21cm,
 require high spectral resolution.
 The largest {untargeted} \hi\ surveys to date 
 rely on single-dish telescopes,
 lacking spatial resolution,
 both to provide sensitivity and 
 backends with the required high spectral resolution over
 relatively large bandwidths, both of which are
 necessary 
 to enable effective
 surveys. 
 The state of the art for
 \hi\ surveys is 
the Arecibo Legacy Fast ALFA  \hi\ survey \citepads[ALFALFA][]{2018ApJ...861...49H},
with the \hi\ Parkes All-Sky Survey \citep[HIPASS][]{2004MNRAS.350.1195M}
providing a complementary view in the 
Southern Hemisphere.

While historically radio continuum and polarization surveys
have been distinct from spectral line surveys, 
the new generation of backends for
interferometric radio telescopes
can handle large bandwidths at high spectral resolution.
The large bandwidth provides for sensitive radio continuum and
polarization data while the high spectral resolution
enables spectral line observations.
  Another key technological advance is the development
 of phased-array feeds (PAFs), which provide a large
 number of beams on the sky, and hence
 large fields of view.
Thus, 
a new generation of interferometers
operating at GHz frequencies
 provide the spatial resolution, sensitivity, and field-of-view
 necessary to efficiently carry out 
 cutting-edge surveys
 both for \hi\ and for radio
 continuum and polarization \citepads{2016MNRAS.460.3419M}.
Examples of this include:
the MeerKAT International
GigaHertz Tiered Extragalactic Exploration
(MIGHTEE) survey with MeerKAT,
originally designed as continuum
survey { but
which is simultaneously a sensitive \hi\ survey }
\citepads{2016mks..confE...6J, 2021A&A...646A..35M};
and  {the synergy between} the surveys on ASKAP, where the
\hi-focused Widefield ASKAP L-band Legacy All-sky
Blind surveY (WALLABY) 
 \citepads{2020Ap&SS.365..118K}
 provides a higher frequency measurement to complement
 the lower frequency
 the radio continuum and polarization surveys, Evolutionary Map of the Universe \citepads[EMU,][]{2021PASA...38...46N} and 
Polarisation Sky Survey of the Universe's Magnetism  \citepads[POSSUM,][]{2010AAS...21547013G}. This synergy between
different observing frequencies can also extend
to looking at spectral lines in both absorption and emission,
such as between WALLABY and the First Large Absorption Survey in \hi\ \citepads[FLASH,][]{2022PASA...39...10A}.



Another example is the PAF
upgrade to the Westerbork Synthesis Radio Telescope (WSRT), the APERture Tile In Focus array (Apertif), fully described in \citet{2022A&A...658A.146V}.
Apertif is a complex system;
here, we briefly summarize the most relevant aspects from
the perspective of an imaging science user.
The Apertif system consists of PAFs with 121 Vivaldi elements on 12 of the 14 WSRT dishes, including dishes RTC and RTD which
provide the longest baselines (and highest angular resolution).
The signal from these elements can be combined to form forty 
simultaneous {beams} on the sky, each with a typical
half-power size of 35\arcmin, significantly increasing the 
instantaneous field of
view of WSRT to {$\sim$8} deg$^2$ at 50\% of the peak sensitivity.
The layout of the beams is shown in figure 25 of \citetads{2022A&A...658A.146V}.
The Apertif system provides 300 MHz of contiguous bandwidth within the frequency range 1130-1750 MHz.
At the dish level, the 300 MHz bandwidth is divided
into 384 subbands of {781.25} kHz each.
The backend can be configured to provide an imaging correlator
which further divides each subband into 64 channels of 12.2 kHz.

Following this upgrade, WSRT-Apertif was dedicated to large
survey programs:
a time-domain survey 
\citepads[][van Leeuwen \etal in prep]{2017ursi.confE...2M}
 and
a two-tiered imaging survey (Hess \etal in prep). 
The legacy of a survey is greatly enhanced
by the release of high quality science-ready
data products to the community.
Toward that end, we present
the first release of Apertif imaging survey data,
covering the first year of survey operations,
1 July 2019 -- 30 June 2020.
In this first data release, the emphasis is on releasing the best quality data products from a first processing
to the community as rapidly as possible to demonstrate the
potential of the Apertif imaging surveys.

The paper is organized as follows:
Section \ref{sec:surveys} introduces the Apertif imaging surveys,
and the observations covered by this release.
Section \ref{sec:data} describes the released data products,
including the choice for which {of these} to release.
The quality of the continuum, polarization, and spectral line
data is discussed in Sections \ref{sec:cont}, \ref{sec:pol},
and \ref{sec:hi}.
In Section \ref{sec:caveats} we detail 
known caveats and limitations of the Apertif system and data, and we strongly encourage the reader to carefully read this section.
We highlight the science potential of the Apertif imaging
surveys in Section \ref{sec:science}.
In Section \ref{sec:future}, we briefly describe future prospects and improvements for further data releases.
Finally, Section \ref{sec:summary} offers a brief summary
of the data release.
{Appendix \ref{app:otherdata} provides information on available data from 
pre-survey operations, including a science verification campaign (SVC)
undertaken to verify the scientific
performance of the Apertif system and early science observations
between the SVC and start of survey operations.}
A companion paper, \citetads[][hereafter K22]{K22}, presents a continuum source
catalog for the first data release.


\section{The Apertif imaging surveys}\label{sec:surveys}





The wide field-of-view of Apertif, 
the {spectral resolution of 12.2 kHz across a large bandwidth of 300 MHz, 
and an angular resolution of up to 11\arcsec/$\sin\delta$}
enable large-scale, simultaneous surveys
of the radio continuum, polarization,
and spectral line {sky}.
Toward this end, a significant fraction of the Apertif observing time
has been dedicated to a legacy survey program consisting of two tiers
of imaging surveys: a wide and medium-deep tier, referred to respectively as the Apertif Wide-area Extragalactic Survey (AWES) and the Apertif Medium-deep Extragalactic Survey (AMES). The wide tier aims to cover a relatively large region of the sky,
while the medium-deep tier focuses on a smaller region of sky, visiting individual fields 
{up to} ten times in order to build up
sensitivity, with the ultimate goal of achieving
a column density sensitivity of $\sim$5$\times10^{19}$ {atoms} cm$^{-2}$ in order to detect low surface brightness \hi. The repeated observations of the medium-deep
tier also enable variability studies.

\begin{figure*}
    \centering
    \includegraphics[width=0.9\linewidth,keepaspectratio]
    {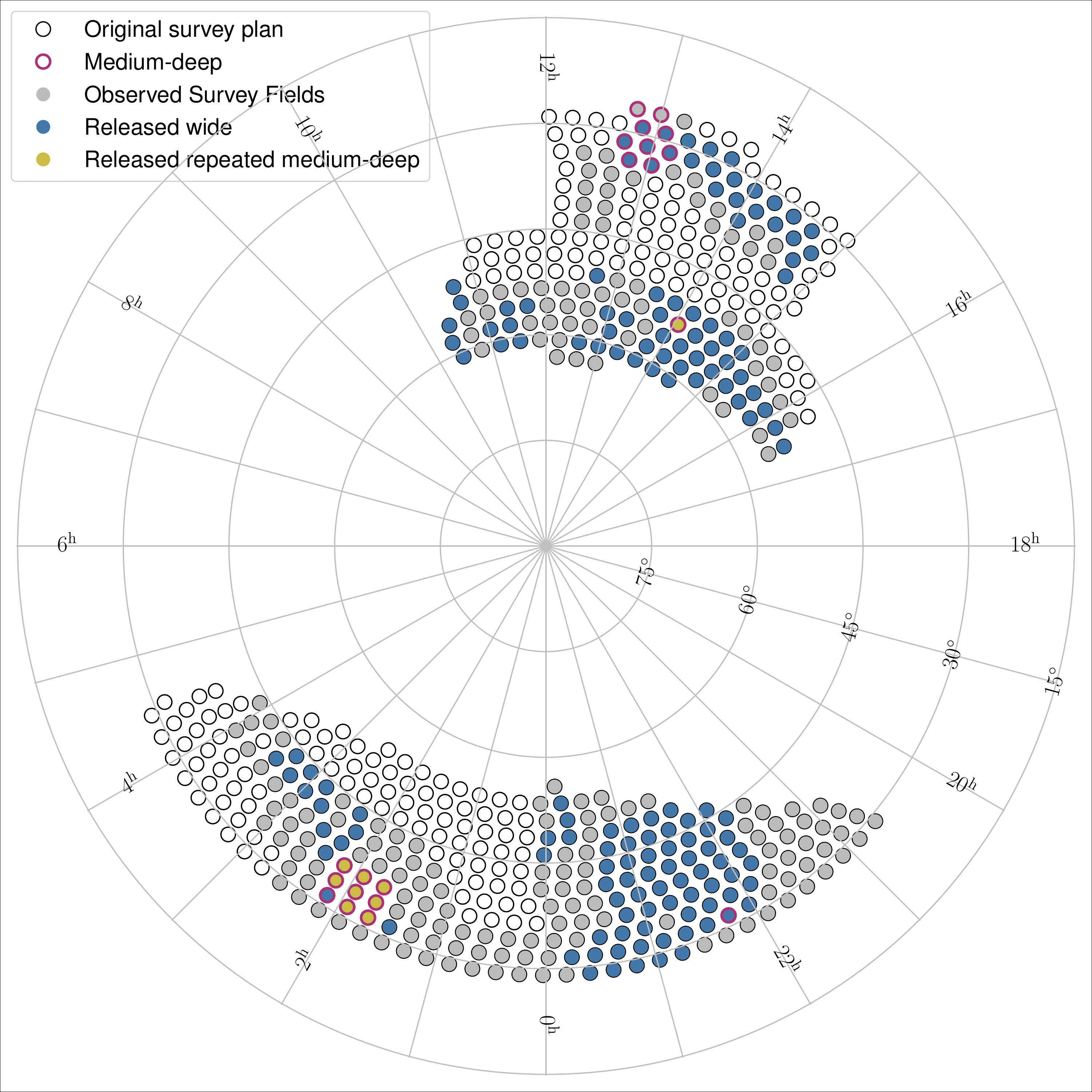}
    \caption{Sky view of Apertif imaging observations. Blue circles indicate
    wide observations (single visit) while pink circles indicate fields with multiple visits as part of this data release (but we note that data products are only released for individual observations). The open black circles indicate
    the coverage in the original four-year survey plan. Grey circles show Apertif observations to date; these effectively represent the final survey footprint. 
    Circles outlined in purple have medium-deep coverage in the full survey.}
    \label{fig:observations}
\end{figure*}

The design of the imaging surveys was driven by a broad set of extragalactic science goals, including:
\begin{itemize}
\item Investigate gas and total dynamical mass distribution (derived from gas kinematics) over
a wide range of galaxy properties, using resolved \hi\ observations
\item Investigate environment and the role of interactions, gas accretion and removal of gas 
\item Investigate the properties of the smallest gas-rich galaxies in the local Universe
\item Study the role of cold gas in active galactic nuclei (AGN) and their feedback activity
\item Study the history of star formation and AGN activity of the faint radio continuum population
\item Study the magnetic fields in galaxies and of the large-scale structures in which they are embedded
\end{itemize}
With these goals in mind, the footprint of the Apertif imaging surveys was selected to maximize multiwavelength coverage, especially optical redshifts to
provide distances for radio continuum sources.
Other considerations included maximizing overlap with LoTSS \citep{2019A&A...622A...1S},
providing a large coverage in Right Ascension,
and maximizing scheduling feasibility and efficiency.
Figure \ref{fig:observations} provides an overview of the 
imaging survey footprint. The full sky coverage
of the original four-year survey plan is shown for reference;
filled circles indicate the effective final coverage of the Apertif
imaging surveys (colored circles are contained within this data release).
Fields that have medium-deep coverage in the full survey
are also highlighted in Figure \ref{fig:observations}. There
are two main regions of interest for the medium-deep tier:
the Perseus-Pisces supercluster, a rich environment, and the Herschel-ATLAS North Galactic Pole field \citep{2017ApJS..233...26S}, which has large amounts of ancillary data available. 
In addition, there {are} two individual fields hosting intra-hour variables
with medium-deep coverage.
Hess \etal (in prep) provide a detailed discussion on the science cases, chosen footprint, and tiling strategy.

During the first year of survey observations, medium-deep
observations were focused on the {Perseus-Pisces} field.
An additional medium-deep field
containing the first intra-hour variable 
found in the Apertif imaging surveys \citepads{2020A&A...641L...4O} was also included.
The wide observations focused on coverage between $10-16h$ R.A. with Decl. $>45$\dg\ to maximize overlap with publicly available LoTSS data. Wide observations in the same R.A. range but at lower Decl. were also carried out to start building coverage around the second medium-deep footprint. The wide tier
in the first year also included significant coverage at  $22-0h$ R.A. for maximizing schedule efficiency.


The two tiers share a common observing strategy and observational setup.
Survey fields are observed for 11.5 hours
as a  compromise between flexibility in scheduling and the depth/maximum uv coverage of a full 12 hour synthesis imaging track\footnote{As WSRT is an east-west array, Earth rotation aperture synthesis is critical for uv coverage.}. 
Calibrator observations alternate between a flux/bandpass (3C147 or 3C196, occasionally 3C295) and polarization calibrator (3C286 or 3C138); calibrators are observed in every
Apertif beam, necessitating forty separate calibrator observations.
The calibrator scans vary from $3-5$ minutes in length {per compound beam}, depending upon the available gap until the next survey field. Up to two survey fields can be observed in a row; thus a survey field may be separated in time from its calibrator
by up to $\sim$15 hours. 
Occasionally an observing session will start or end with a survey field in order to increase observing efficiency; then the separation in time may be longer.
In order to validate the long time separation between calibrators and target fields, we checked calibration gain solutions over
the course of a full observing run (10 days). The {amplitude}
solutions
varied by less than 10\% over the full period,
demonstrating that {a} longer period between calibration
scans should have minimal impact on the absolute calibration.
Hess \etal (in prep) describes the automation and details of the survey scheduling.
As the main spectral line of interest is \hi,
the observations use the lowest frequency settings,
covering the band 1130-1430 MHz, corresponding to a
{maximum} redshift of 0.26 for \hi.
This band offers a large cosmic volume for observing \hi;
however, the lower half of the band is significantly affected by RFI
and thus no derived data products are offered for it (see Section \ref{sec:released_proc_data})\footnote{We note that
for survey observations starting from February 2021 the observing
band changes to 1220-1520 MHz to provide the largest
RFI-free band possible.}.
The 24576 channels provide a frequency resolution of 12.2 kHz,
or 2.6--3.2  \kms\ across the band for \hi,
{but spectral line data products are typically
produced with a spectral binning to 36.6 kHz}.
The angular resolution of the imaging survey observations
is 11--15\arcsec$\times$11--15\arcsec/sin($\delta$),
{as the
exact resolution depends on the weighting used. Continuum and polarization data are imaged with uniform weighting to provide the
highest angular
resolution, close to 11\arcsec, while the line cubes
are produced with a resolution of $\sim$15\arcsec\ and increased
surface brightness sensitivity.
We emphasize that}
the two outlying dishes, RTC and RTD, are required for full angular
resolution. The achieved noise in a single observation is 
$\sim$40 $\mu$Jy \bm\ in the continuum images and $\sim$1.6 mJy \bm\ over 36.6 kHz in the line cubes.




\section{Data release}\label{sec:data}

This section describes the released Apertif data products.
All data from the first year of survey observations are considered for release. This includes all
raw observational data, and a selection of processed data.
In order to balance data quality against the timeliness of a prompt release,
processed data products are only released if the continuum image
passes validation as described in Section \ref{sec:cont_qa}.
This decision is further motivated below in Section \ref{sec:released_proc_data}.
An overview, including the original size, of all Apertif data products is provided in Table \ref{tab:dataproducts}.

\begin{table*}[]
\caption{List of available data products}
\label{tab:dataproducts}
\centering
\begin{tabular}{lllll}
\hline
\hline
Data product & Format & 
Dimensions\tablefootmark{a} & 
Size & Size \\ 
 & & & per beam & per observation \\
\hline
    \multicolumn{4}{l}{\textit{Level zero data products}} \\
    Survey field raw visibility data &  MS & 
        $4\times24576$ & 
        117 GB &  4.7 TB \\
    Calibrator raw visibility data & MS & $4\times24576$
    & $1.6-2.6$ GB\tablefootmark{b} & $64-104$ GB\tablefootmark{b}\\
    \hline
    \multicolumn{4}{l}{\textit{Level one data products}} \\
    Calibration tables & MS table & 
        -- 
        & 660 MB & 26.3GB \\
    Self-calibrated visibility data  & uvfits & $4\times12288$
    & 58 GB &  2.3TB\tablefootmark{c}\\
    \hline
    \multicolumn{4}{l}{\textit{Level two data products}} \\ 
    \multicolumn{4}{l}{\textit{Continuum}} \\ 
    Multi-frequency synthesis beam images & fits & 3073$\times$3073 & 37 MB & 1.5 GB\\
    \multicolumn{4}{l}{\textit{Polarization}} \\ 
     Stokes Q and U cubes   & fits & $2049\times2049\times24$ & 1.5 GB & 62 GB\\
    Stokes V multi-frequency synthesis image  & fits & $3073\times3073$ & 37 MB & 1.5 GB\\
     \multicolumn{4}{l}{\textit{Line}} \\ 
     Continuum-subtracted, dirty line cubes   & fits & $661\times661\times1218$ & 8 GB\tablefootmark{d} & 320 GB\tablefootmark{d}\\
    Restoring beam cubes   & fits & $661\times1321\times1218$ & 320 GB\tablefootmark{d,e} & 640 GB\tablefootmark{d,e}\\
     
    \hline
\end{tabular}
\tablefoot{
\tablefootmark{a}{For level zero and one data: number of polarizations and channels. For level two data: {i}mage/cube sizes.}
\tablefootmark{b}{Data sizes are after pruning, to only keep the Apertif beam that contains that calibrator{, and the range} of values is for different calibrator {scan} lengths.}
\tablefootmark{c}{Maximum size; files
can be smaller depending on flagging within pipeline.}
\tablefootmark{d}{Total size for four line/beam cubes produced.}
\tablefootmark{e}{After symmetric reduction.}
}
\end{table*}


\subsection{Apercal processing}

Survey observations are automatically
processed by {\tt Apercal}\footnote{Publicly available at \url{https://github.com/apertif/apercal}}, the Apertif imaging and calibration pipeline, via an automated trigger, {\tt autocal}. Full details of this pipeline are provided in 
\citetads{2022A&C....3800514A}. 
A brief description of the processing is provided below.


The {\tt Apercal} pipeline is a modular pipeline written in python using common astronomical software.
Apercal processes each beam of an Apertif observation individually, following standard reduction procedures. 
First, the data is split to only keep frequency ranges mostly free
of radio frequency interference (RFI). Experience showed that
removing frequency ranges dominated by RFI significantly improved
the calibration and quality of the data at all frequencies.
For the data under consideration for the data release,
this meant splitting to keep the upper 150 MHz of the band,
1280-1430 MHz, and additionally flagging the first 12.5 MHz, in
order to avoid persistent RFI, resulting in an effective frequency range of $1292.5-1430$ MHz.
The data is then flagged, both for a-priori known 
flags\footnote{It should be noted that {while {\tt Apercal}
supports robust, unique flagging
commands for any observation, the automated processing pipeline workflow ({\tt autocal}) uses a standard set of 
flags for all observations. Thus it does not provide any additional flags due to known but temporary issues,
such as stuck dishes that were not on source. When these issues occasionally arise, a manual processing is undertaken
where appropriate flag commands are provided.
}} and for RFI using {\tt AOFlagger}, including special
adaptations for Apertif \citepads{aoflagger}.
A standard cross-calibration (including of cross-hand data products)
is then done using {\tt CASA} tasks \citepads{2007ASPC..376..127M}.
A direction-independent self-calibration is  undertaken
in {\tt miriad} \citepads{1995ASPC...77..433S} on data frequency-averaged to the subband resolution.
{As a first step, a phase-only parametric self-calibration based on FIRST
and NVSS is undertaken; then the self-calibration iterations
on the data are undertaken. }
Phase solutions are always found {and applied to the data; amplitude solutions are always derived but only applied to the
data if the subsequent dirty image with amplitude solutions improves on the dirty image with only
phase solutions.}
Importantly, Apercal does not account for any direction dependent errors.
Finally,
the self-calibration solutions are applied and the
imaging of the continuum, polarization and line data is done in {\tt miriad}. For the continuum images, sources above 5-$\sigma$ 
{are identified}
and cleaned
to a threshold of 1-$\sigma$,
where $\sigma$ is the ``theoretical'' noise, calculated
as the noise in a Stokes V image.
The same masks are applied to the polarization data and
sources there are also cleaned to a 1-$\sigma$ threshold.
The model of the continuum emission is subtracted
from the full spectral resolution data and line cubes are created
\footnote{
{As discussed in Section \ref{sec:hi_qual},
over 75\% of released cubes show no (or extremely minor)
artifacts; of the remainder, 15\% show significant artifacts.
These artifacts include incomplete continuum subtraction,
in addition to other data quality issues. Thus the single
step of visibility-based continuum subtraction is sufficient in the majority of cases.
}}.
We note that the line cubes are {only} imaged, meaning
that  emission is {neither} identified
{nor} cleaned in them; {for this reason
the dirty beam cubes are also kept as a data
product to enable later cleaning}.
Finally, the processed data products are automatically ingested back into the Apertif Long-Term Archive (ALTA).


\subsection{Virtual Observatory interface}
The released data products are exposed through standard Virtual Observatory (VO) protocols to facilitate their access and exploration to both general and specialized users.
In particular, the protocols offered are the Tabular Access Protocol (TAP)\footnote{http://www.ivoa.net/documents/TAP/} and the Simple Image Access protocol (SIA)\footnote{http://www.ivoa.net/documents/SIA/}.

The VO registry can be accessed also through {a} web browser\footnote{At the address \url{https://vo.astron.nl}}, which presents
all collections on the ASTRON VO registry, 
including APERTIF\_DR1. Selecting one collection allows the user to interactively query through a form all data products of that given collection, which is sorted by data product type for Apertif DR1.

The above query capabilities are also possible through standard VO tools, such as using TOPCAT \citepads{2005ASPC..347...29T} via TAP. 
The table names to query the different DR1 collections are summarized in Table \ref{tab:vo_obscore_table}. Moreover, it is possible to query all the available dataproducts from Apertif DR1 at once by using the table \emph{ivoa.obscore} 
and limiting the results to the apertif\_dr1 by appending to the ADQL statement "{\tt where obs\_collection="apertif\_dr1"}".

\begin{table*}
\centering
\caption{VO tables for Apertif DR1}
\label{tab:vo_obscore_table}
\begin{tabular}{llll}
\hline 
\hline
Table name & url\tablefootmark{a} & obscore type & obscore subtype \\ 

\hline 
 \multicolumn{4}{l}{\textit{Level two data products}} \\
 
apertif\_dr1.continuum\_image & 
    \href{https://vo.astron.nl/apertif_dr1/q/apertif_dr1_continuum_images/form}{/apertif\_dr1\_continuum\_images/form} & 
    image & continuum \\ 
apertif\_dr1.pol\_cubes & 
    \href{https://vo.astron.nl/apertif_dr1/q/apertif_dr1_polarization_cubes/form}{/apertif\_dr1\_polarization\_cubes/form} & 
    cube & polarization cube \\ 
apertif\_dr1.spectral\_cubes & 
    \href{https://vo.astron.nl/apertif_dr1/q/apertif_dr1_spectral_cubes/form}{/apertif\_dr1\_spectral\_cubes/form} & 
    cube & spectral cube \\ 
apertif\_dr1.beam\_cubes & 
    \href{https://vo.astron.nl/apertif_dr1/q/apertif_dr1_spectral_cubes/form}{...} & 
    cube & dirty beam \\
    \hline 
 \multicolumn{4}{l}{\textit{Level one data products}} \\
 apertif\_dr1.calibrated\_visibilities &
    \href{https://vo.astron.nl/apertif_dr1/q/apertif_dr1_calibrated_visibilities/form}{/apertif\_dr1\_calibrated\_visibilities/form} & 
    visibility & calibrated visibility \\
    
\multicolumn{4}{l}{\textit{Level zero data products}} \\
apertif\_dr1.raw\_visibilities & 
    \href{https://vo.astron.nl/apertif_dr1/q/apertif_dr1_raw_visibilities/form}{/apertif\_dr1\_raw\_visibilities/form} & 
    visibility & raw visibility \\
apertif\_dr1.flux\_cal\_visibilities & 
    \href{https://vo.astron.nl/apertif_dr1/q/apertif_dr1_flux_cal_visibilities/form}{/apertif\_dr1\_flux\_cal\_visibilities/form} &
    visibility & raw visibility \\
apertif\_dr1.pol\_cal\_visibilities & 
    \href{https://vo.astron.nl/apertif_dr1/q/apertif_dr1_pol_cal_visibilities/form}{/apertif\_dr1\_pol\_cal\_visibilities/form} &
    visibility & raw visibility \\
\hline
\end{tabular} 
\tablefoot{
\tablefootmark{a}{The URL are expressed relative to \url{https://vo.astron.nl/apertif_dr1/q}}
}

\end{table*}

\subsection{Extent of data release}


All raw observational data from the first year of survey operations are released.
A subset of processed data products, meeting
specified quality assessment criteria, are also released.
These two subsets of released data are discussed further below.

\subsubsection{Raw observational  data}
From the first year of survey observations, there are
221 observations of 160 unique survey fields; with an effective field-of-view of 6.44 square degrees\footnote{Based on a spacing between fields of 2.3\dg\ in Dec. and 2.8\dg\ in R.A.}, these observations cover over 1000 square degrees of sky.
There are eight medium-deep fields with multiple observations within this data release
(64 observations in total, not evenly distributed among the fields); these fields
are highlighted in Figure \ref{fig:observations}.

The raw  observational data is recorded in measurement-set  (MS) format. The sizes of the original raw data are given in Table \ref{tab:dataproducts}. A set of calibration scans consists of 40 separate observations taken in succession. For each calibrator scan only one beam contains the calibrator; all other beams not containing the calibrator are discarded to save on data volume. 
We note that the calibrator observations are taken at a higher time resolution than the survey fields (10 vs. 30 seconds) to allow better RFI excision due to their shorter integration time.

As can be seen in Table \ref{tab:vo_obscore_table}, the raw visibility data
are provided as separate collections for the survey fields, flux calibrator 
and polarization calibrator observations. The tables
can be joined to identify the associated calibrators
for a target observation.

The raw observational data are not stored on disk but rather on tape at a national facility, SURFsara\footnote{\url{https://userinfo.surfsara.nl/}}. In addition, the raw data are currently being compressed to further save on data volume
\citepads{2016A&A...595A..99O}. Thus,
the VO tools enable
queries of these data products, but not direct access or download. 
The data release documentation\footnote{\url{http://hdl.handle.net/21.12136/B014022C-978B-40F6-96C6-1A3B1F4A3DB0}} contains up-to-date information on how to access the data.



\subsubsection{Processed data}\label{sec:released_proc_data}

Processed data products that can be used directly,
or with minimum additional processing,
are of the most interest
to the broader scientific community.
The choice to release processed data products is done
on a beam-by-beam basis, based on the quality of the 
continuum image. The continuum data is used for the self-calibration, so an artifact-free continuum image is a
good indication that the inherent data quality {of the observation} and {the} calibration
are of a sufficient standard. This is also a practical choice
as the structure of the ingested processed data in ALTA means
that either all or no processed data for a given beam should be released. Thus, in order for processed data products for a given beam to be released, the continuum image must pass
validation. 
This approach offers a compromise between
releasing as much data as possible on a short timescale to
the community while only releasing
data that is of high enough quality to be of sufficient interest to the community.

{Toward this end, we define quality metrics for continuum images, polarization
images and cubes, and spectral line cubes in Sections
\ref{sec:cont_qa}, \ref{sec:pol_qa} and \ref{sec:hi_valid}.
The metrics are focused on ensuring a minimum sensitivity and resolution, in addition to identifying significant artifacts in images. The continuum validation criteria in Section \ref{sec:cont_qa} must be satisfied for processed data products to be released. The criteria for polarization and spectral line data products do not need to be satisfied, but the data are released with clear flags that indicate if they fail; the frequency of this
is discussed in Sections \ref{sec:pol_valid} and \ref{sec:hi_qual}. We additionally provide the metrics for all data products
in the relevant VO tables.
}

A total of 7640 beams were considered for release; 3374 of these
beams are included in this first data release.
We note that the unreleased beams will be contained in future data releases with improved processing (see Section \ref{sec:future}).
The sky coverage of the released continuum images
is available as a HiPS image\footnote{Via \url{https://hips.astron.nl/ASTRON/P/apertif_dr1/}}.

There are two additional important notes that are relevant for these data products. First, processed data products are only available for the upper 150 MHz of the band; thus the processed data products are produced over the range 1280-1430 MHz. Moreover, the first 12.5 MHz of data are flagged due to persistent RFI, therefore the resulting central frequency is 1361.25 MHz. The nominal bandwidth is then 137.5 MHz, but effectively it could be smaller due to additional RFI flagging.
Secondly, these data products are not primary-beam corrected; primary-beam images are provided separately (see Section \ref{sec:data_pb}); these may be used for mosaicking or for correction of individual images.

The available processed data products are briefly discussed
below. The quality of the continuum images, polarization images and cubes, and spectral line
cubes
are discussed in more detail in Sections \ref{sec:cont},
\ref{sec:pol}, and \ref{sec:hi}.

\begin{itemize}

\item \textbf{Calibrated visibility data}
Calibrated visibility data, with
cross-calibration and self-calibration solutions applied,
are currently stored as an intermediate
data product at full time and spectral resolution; this may change with future updates to save storage space. These are stored on tape with the raw data, and the data release documentation\footnote{\url{http://hdl.handle.net/21.12136/B014022C-978B-40F6-96C6-1A3B1F4A3DB0}} should be consulted for access.

\item \textbf{Continuum images}
A multi-frequency synthesis (mfs) Stokes I image is created over the full  frequency range (1292.5 -- 1430 MHz) and saved as a fits
file for each beam. The size
of the continuum images is 3.4\dg$\times$3.4\dg (3073$\times$3073 pixels, with 4\arcsec/pixel). This extends  well outside
the primary beam response to account for cases
where a strong source is in a sidelobe and needs
to be included in the self-calibration model and cleaning.
Sources at the 5-$\sigma$ level should be identified and cleaned
to the 1$-\sigma$ level\footnote{
Because the Apercal masking procedure uses thresholds that also depend on the maximum value of the residual image
\citep{2022A&C....3800514A}, if there are persistent artifacts in a residual image, the maximum number of clean cycles may be reached before sources down to the 5-$\sigma$ level are included in the source mask.}.

\item \textbf{Polarization images and cubes}
 A mfs Stokes V image over the full bandwidth (1292.5 -- 1430 MHz) is produced.
 This image matches the continuum image in spatial extent:
 3.4\dg$\times$3.4\dg (3073$\times$3073 pixels, with 4\arcsec/pixel).
 In order to prevent bandwidth depolarization and enable
rotation measure synthesis studies, Stokes Q and U cubes with
a frequency resolution of 6.25 MHz are produced.
The cubes have a smaller spatial extent of 2.7\dg$\times$2.7\dg
(2049$\times$2049 pixels, with 4\arcsec/pixel) but still
extend well beyond the primary beam.

\item \textbf{Line and dirty beam cubes}
Four line cubes over a set of different frequency ranges
are produced. Table \ref{tab:lineparams} summarizes
the covered frequency  ranges and provides
the corresponding redshift range for \hi. 
The lowest redshift cube is produced at full spectral resolution while other cubes are produced with a 3-channel averaging. 
These cubes have a spatial extent of 1.1\dg$\times$1.1\dg 
(661$\times$661 pixels, with 6\arcsec/pixel).
As the Apercal pipeline does not provide source finding or cleaning of the line cubes, corresponding dirty  beam cubes 
are also provided to allow offline cleaning of source emission.
Originally these dirty beam cubes are produced with twice the spatial coverage of the line cubes
to enable full cleaning\footnote{We take advantage of the symmetric nature
of the WSRT restoring beam to trim the beam cubes in half around the 
North-South axis to halve the size of the beam cubes.
A script and instructions for restoring the full beams
cubes are available at \url{https://github.com/apertif/trim_apertif_beam_cube}}.
These dirty beams cubes are provided as linked data products in the spectral cube VO table.

\end{itemize}

\begin{table*}[]
    \caption{\hi\ data cube parameters}
    \label{tab:lineparams}
    \centering
    \begin{tabular}{lllrll}
    \hline \hline
    Cube & Frequency Range & 
    Frequency Resolution & 
    Redshift range &
    Velocity Range\tablefootmark{a} & 
    Velocity Resolution\tablefootmark{a,b} \\
     & MHz & \kms & & kHz & \kms \\
    \hline
       Cube0 & 1292.5 -- 1337.1 & 36.6 & 0.062 -- 0.099 & 18678 -- 29667 &  9.0 \\
       Cube1 & 1333.1 -- 1377.7 & 36.6 &  0.031 -- 0.065 & 9293 -- 19634 &  8.5 \\
       Cube2 & 1373.8 -- 1418.4 &  36.6 & 0.001 -- 0.034 & 424 -- 10170 &   8.0 \\
       Cube3 & 1414.5 -- 1429.3 & 12.2 & {-0.006}     -- 0.004 & -1865 -- 1252  &   2.6 \\
    \hline
    \end{tabular}
    \tablefoot{
    \tablefootmark{a}{Optical velocity definition, with v = cz} \,
    \tablefootmark{b}{For center of cube}
    }
\end{table*}


\subsection{Related data products: Primary beam images}\label{sec:data_pb}
Important additional data products are the primary beam images
for each of the forty Apertif beams. Currently, Apercal does not provide primary beam corrections to any data products. 
Instead, primary beam corrections are applied separately, which
requires primary beam images. Due to the nature of the phased
array feed, each Apertif compound beam has a unique primary beam response
and thus primary beam images are required for each of the forty beams.
{ As Apertif does not offer a holography mode, we undertake
two different {approaches} for measuring the primary
beam response: drift scans across extremely bright sources {\citepads{2022arXiv220509662D}} and reconstruction of the primary beam responses via comparison to the NVSS catalog with a Gaussian process regression technique (K22).
For this first data release, the primary beams from the Gaussian
process regression approach are released. K22
describes their derivation in detail, and the data release
documentation contains information on how
to best apply these primary beam images to the Apertif data, including how to scale the primary beam images with frequency.
}



\section{Continuum image quality}\label{sec:cont}

This section describes the validation of the continuum images
and provides an overview of the data quality.
In addition, we report on the accuracy of the flux scale
and astrometric precision.

\subsection{Validation of continuum images}\label{sec:cont_qa}

The continuum images of every beam are individually validated.
The starting point of the validation is the residual images obtained after cleaning the continuum images, and the validation aims at checking to what extent these images  contain
only Gaussian noise. The premise is that any significant deviation from  
Gaussian noise in the residual image indicates issues with the calibration and reduction of the data.

The following parameters were derived for each residual image: 
\begin{itemize}
    \item $\sigma_{\rm in}$: Noise in the inner half degree of the image\footnote{{Specifically, within the central 500$\times$500 pixels}}. 
    \item $\sigma_{\rm out}$: Noise at the edge of the residual image, more than a degree from the centre. 
    \item $b_{min}$: The size of the minor axis of the restoring beam.
    \item $R=\sigma_{\rm in}/\sigma_{\rm out}$ 
    \item Ex$-$2:  Area, in units of beam area, with {values  $<-2 \sigma$ in }the inner 0.5\dg\ of the 
    residual image, in excess of what is expected from a purely Gaussian distribution. 
    \item {Neg10: the absolute value of the level (in units of $\sigma_{out}$) where
    pixels with values below this level cover 10 beam areas within the inner 0.5\dg\ of the residual image. }
    
\end{itemize}

\begin{figure*}
    \centering
    \includegraphics[width=0.95\linewidth]{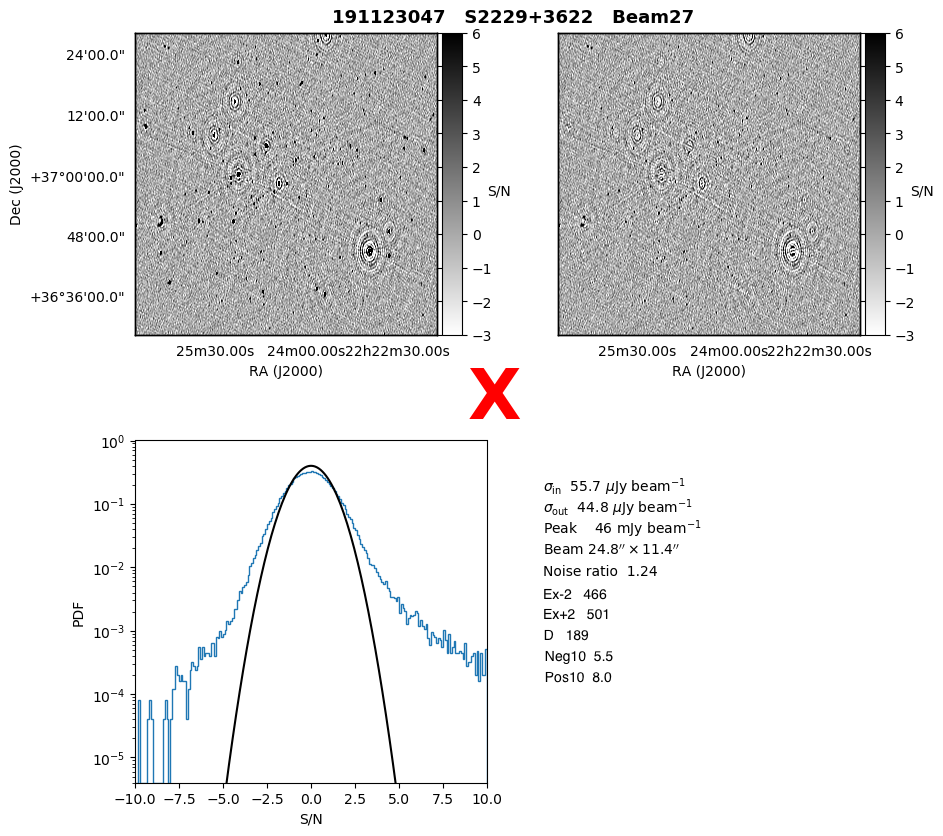}
    \caption{{Example of a continuum image that fails validation
    (as indicated by the red x); beam 27 of ObsID 191123047. The four panels shown are}: \textit{Upper left:} Continuum image;  \textit{Upper right:} Residual image {(note the uncleaned source emission)};
    \textit{Lower left:} {Flux density histogram} of all pixels in the residual image (blue) compared to a Gaussian distribution (black) {corresponding to $\sigma_{out}$}; \textit{Lower right:} Statistics calculated for the {beam. These include the metrics
    in Section \ref{sec:cont_qa}, in addition
    to other statistics not (currently) used in the validation.} {The first five are: $\sigma_{in}$, 
    $\sigma_{out}$, the peak
    in the continuum image, the restoring beam, and 
    $R$. The next two are Ex$-$2 and Ex$+$2. The dynamic range (D) is calculated as the peak of
    the continuum image divided by Neg10$\times \sigma_{out}$. The final two statistics are Neg10 and Pos10.
    }
    }
    \label{fig:cont_valid_fail}
\end{figure*}

\begin{figure*}
    \centering
    \includegraphics[width=0.95\linewidth]{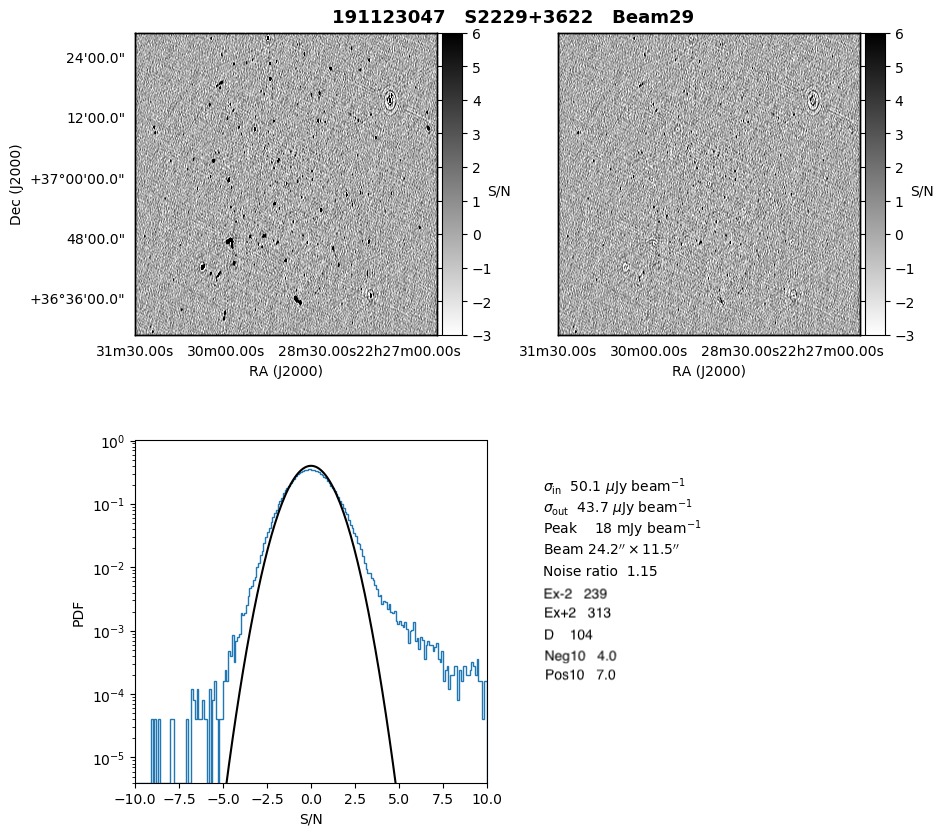}
    \caption{{Example of a continuum image that passes validation; beam 29 of ObsID 191123047. The panels are as in
    Figure \ref{fig:cont_valid_fail}.}}
    \label{fig:cont_valid_pass}
\end{figure*}

{
Both $\sigma_{in}$ and $\sigma_{out}$ are determined in a
robust way using the median of the absolute values.
$\sigma_{in}$ is the relevant noise for scientific analysis,
including the contributions of imaging artifacts,
while $\sigma_{out}$ is a reasonable measure of the expected noise.
$b_{min}$ corresponds to the angular resolution in an image;
it is chosen over the corresponding size of the major axis of the restoring beam ($b_{max}$) as $b_{min}$ depends only on the baselines available
for an observation, while $b_{max}$ depends 
also on the declination of an observation.
$R$ is a measure of the strength of artifacts left in the
center of the residual image.
Ex$-$2 is a measure of how much a residual image deviates
from perfect Gaussian noise; in the case of perfect noise
Ex$-$2$=$0.
This metric is derived by comparing to a Gaussian corresponding
to $\sigma_{out}$.
Figures \ref{fig:cont_valid_fail} and \ref{fig:cont_valid_pass}
demonstrate
the actual noise histograms and comparison Gaussian distributions.
Neg10 is another measure of how much a residual image
deviates from Gaussianity. We emphasize that it {is} defined as an absolute level, so the expected value for a pure Gaussian distribution is 3.2, and more positive values indicate significant (negative) calibration residuals. }
We note that we did not use the  equivalent of the parameter Ex-2 {or Neg10} 
based on positive deviations from Gaussianity (Ex+2, {Pos10}). 
This is because {insufficient cleaning results in 
source emission remaining in the residual image}
which would then dominate the validation {(see the upper
right panel of Figures
\ref{fig:cont_valid_fail} and \ref{fig:cont_valid_pass})}.


{
We then set the following criteria to pass validation; in order
for a continuum image to pass validation  and be released
its residual image
must satisfy all four of the following requirements:}
\begin{enumerate}
    \item  $\sigma_{in}$ $< 60$ $\mu$Jy/beam\footnote{
    Since $\sigma_{in}$ is always greater than $\sigma_{out}$
    this is also a criterion on $\sigma_{out}$
    }. 
    \item  $b_{min}<$15\arcsec. 
    \item $R < 1.225$. 
    \item $R < 1.15$ {or} Neg10 $< 4.5$ {or} $Ex-2 < 400$. 
\end{enumerate}

{
The first two criteria were chosen to ensure {that} minimum survey
specifications were met.
Criterion 1 ensures that the noise of the images is low enough to
be considered survey quality and valid.
Criterion 2 ensures a minimum angular resolution that
is required to achieve the science goals and meet survey specification. The restoring beam can be larger when both dishes with the
longest baselines (RTC and RTD)  are missing from an observation, and observations that are missing both RTC and RTD are considered failed\footnote{
Such observations are repeated for individual wide fields while for medium-deep fields the eventual combination with data that contain RTC and RTD provides the final required angular resolution
}.
}

{
The second two criteria were set to ensure a minimum of imaging
artifacts in continuum images. These criteria were determined by undertaking visual examination of a large set of images
to determine}
 numerical criteria that would catch significant image artifacts. The main types
of image artifacts are due to errors in the self-calibration as well as strong direction-dependent errors for which the current calibration pipeline does not attempt to correct.
{
Criterion 3 ensures that images do not have too much
extra noise in the inner parts of the residual image, corresponding to 
left-over artifacts. Criterion 4 focuses on edges cases,
where $R$ is at intermediate values and additional constraints
on large-scale structures and deviation from Gaussianity in the 
residual image are needed.
We note that the first criterion can also eliminate images
with significant artifacts if these increase the noise
in the inner part of the image above the threshold.}
The criteria were set so that the large majority of images which were visually
assessed as good would pass while 
minimizing the number of images visually assessed as bad which would
{pass.}

Figures \ref{fig:cont_valid_fail} and \ref{fig:cont_valid_pass}
provide an example of the diagnostic plots and
validation metrics for two beams, one of which fails validation
and one which passes. {The image that fails validation
has $\sigma_{in} = 55.7$ $\mu$Jy \bm, $b_{min} = 11.4$\arcsec,
$R=1.24$, Ex$-2 = 466$, and Neg10 $=$ 5.5; thus is fails
both criteria 3 and 4.}
The validation metrics are provided for all released
continuum images in the VO table. 


\subsection{Overview of released continuum image quality}

\begin{figure}
    \centering
    \includegraphics[width=0.9\linewidth,keepaspectratio]{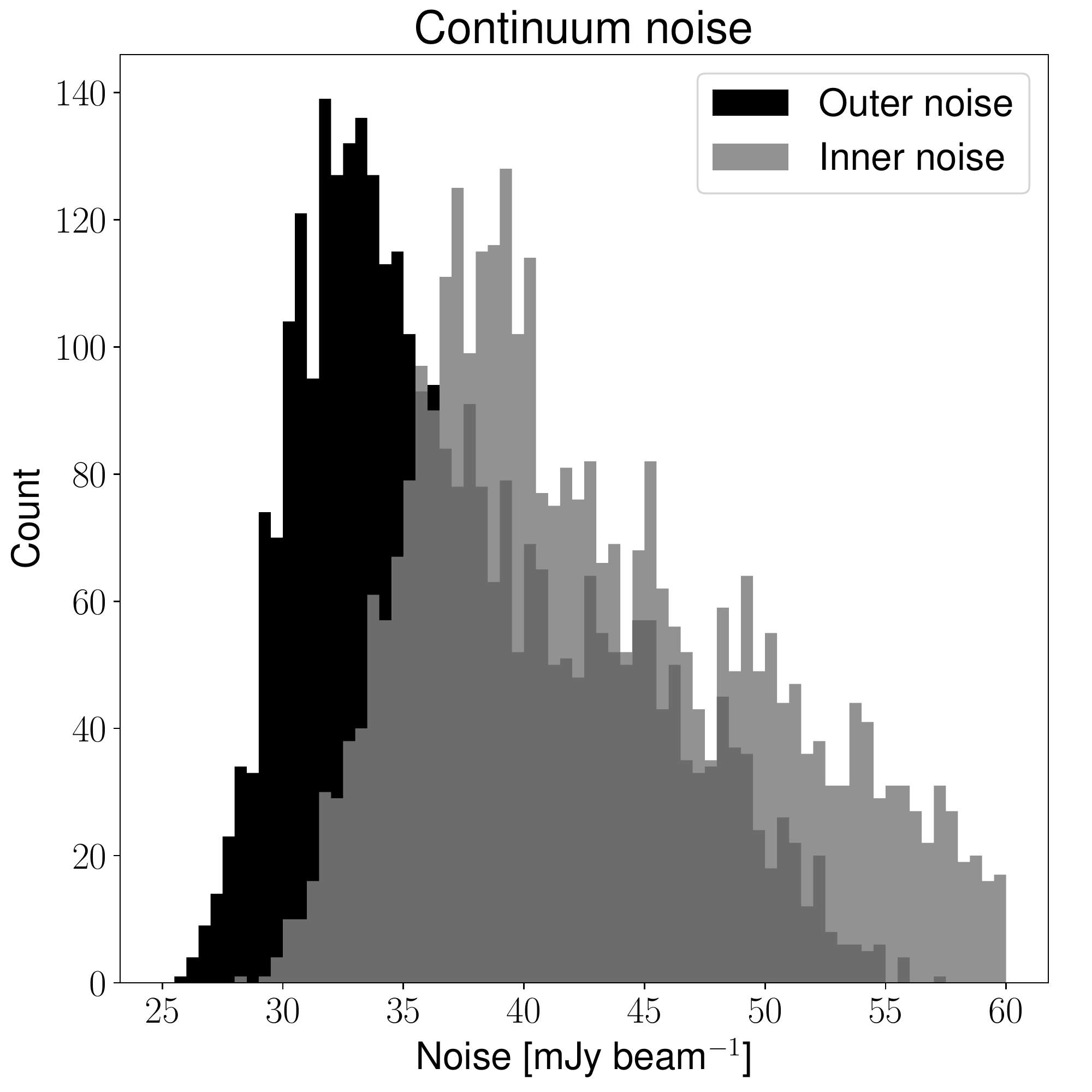}
    \caption{Distribution of inner and outer noise for released continuum images.}
    \label{fig:cont_qual}
\end{figure}

Figure \ref{fig:cont_qual} provides an overview of the
inner and outer noise values for all released (passed validation)
continuum images. 
The median inner noise across all images, relevant for scientific analysis,
is 41.4 uJy \bm. The median outer noise,
indicative of what can be achieved with improved processing,
is 36.1 uJy \bm. These values, along with {uncertainties}
based on the 16$^{th}$ and 84$^{th}$ percentile ranges, are presented in Table \ref{tab:noise}.

\begin{table*} 
\centering 
\caption{Typical noise values {of released data products}} 
\label{tab:noise} 
\renewcommand{\arraystretch}{1.2} 
\begin{tabular}{lll} 
\hline \hline 
Data product & Released\tablefootmark{a} & Passed\tablefootmark{b} \\ 
\hline 
Continuum {image, inner region} ($\mu$Jy \bm) & $41.4^{+9.5}_{-5.4}$  & -- \\ 
 Continuum {image, outer region} ($\mu$Jy \bm) & $36.1^{+8.7}_{-4.9}$ & -- \\ 
 Stokes V {image, inner region} ($\mu$Jy \bm) & $36.9^{+9.2}_{-4.8}$  & $36.5^{+8.9}_{-4.4}$  \\ 
 Stokes V {image, outer region} ($\mu$Jy \bm) & $35.8^{+8.9}_{-4.5}$  & $35.6^{+8.7}_{-4.3}$  \\ 
 Cube0 (mJy \bm ) & $1.66^{+0.33}_{-0.19}$  & $1.64^{+0.31}_{-0.18}$  \\ 
 Cube1 (mJy \bm ) & $1.62^{+0.36}_{-0.19}$ & $1.59^{+0.31}_{-0.17}$ \\ 
 Cube2 (mJy \bm )& $1.60^{+0.35}_{-0.19}$ & $1.56^{+0.32}_{-0.16}$  \\ 
 \hline 
\end{tabular}
\tablefoot{ 
{\tablefootmark{a}{For all released beams, 
which is based on the continuum image passing validation}
\tablefootmark{b}{For only those beams which pass validation for 
the specified data product (good quality for line cubes,
see Section \ref{sec:hi_valid});
for continuum images this is the same as the released beams, while it is a subset for polarization and line. } }
}
\end{table*}

\begin{figure}
    \centering
    \includegraphics[width=0.9\linewidth]{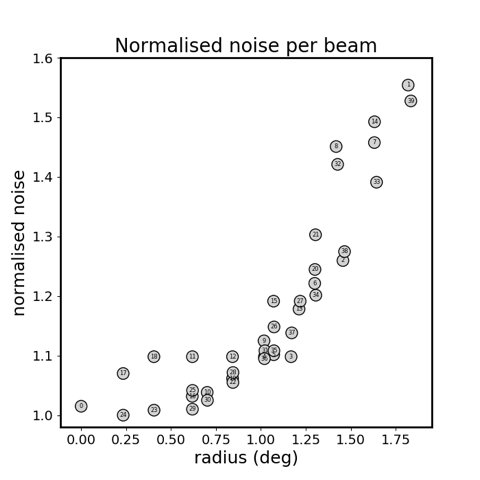}
    \caption{Noise of continuum images for different Apertif beams,
    {relative to the beam with the lowest noise (beam 24), }as a function of distance from the pointing center. {After mosaicking the noise is flatter.}
    }
    \label{fig:beam_noise_dist}
\end{figure}

The behavior of different compound beams is not identical. Specifically, the outer compound beams illuminate the edge of the field of view and thus may be expected to have a reduced sensitivity. This is demonstrated
in figure 38 of \citetads{2022A&A...658A.146V} which shows the 
noise of continuum images in the compound beam layout,
{normalized to the beam with the lowest average noise,
which is beam 24}. Beams closer
to the edge of the field of view have higher continuum noise values.
Figure \ref{fig:beam_noise_dist} quantifies this by showing the normalized noise as a function of distance from the pointing center of the PAF; the increased noise values track with distance. {We note that mosaicking beams flattens the noise
and reduces this effect.}



\subsection{Accuracy of flux scale}\label{sec:flux}

{K22 assess the accuracy of the flux scale.
They cross match the Apertif continuum catalog to the NVSS catalog and find that the flux scales
match, which is by design of the primary beam derivation. They find a significant
scatter in the flux ratios, which is a combination of multiple effects,
including long-term variability of sources and the different angular
resolution and sensitivities of the two catalogs.
In order to assess the precision of the Apertif flux measurements,
K22 compare flux measurements for medium-deep epochs with
at least five observations,
{and find that flux variations across time indicate that the measurement precision is better than 10\%,}
consistent
with expected errors from cross-calibration.}


\subsection{Astrometric accuracy}\label{sec:astrometry}
{
K22 perform a cross-match
of the Apertif radio continuum catalog to
the public LoTSS catalog \citepads{2022A&A...659A...1S}.
As part of this cross-match they confirm the astrometry
of the Apertif radio continuum images, finding
{that} the typical offset between the Apertif and LoTSS sources is
0\arcsec\ with a standard deviation of 2\arcsec\ in both
the R.A. and Decl. directions.
}

We note that the self-calibration of the Apertif data begins with a parametric 
self-calibration using {FIRST} or NVSS data; thus the Apertif data
is calibrated to the {FIRST} and NVSS astrometry by design.




\section{Polarization data quality}\label{sec:pol}

Compared to the continuum 
images, there are a few important points to keep in mind for the polarization data quality. 
 First, due to the physical nature of Stokes Q, U and V, {astrophysical} emission in the polarization images and cubes can be a positive or negative quantity while it can only be positive in Stokes I. Secondly, for Stokes Q and U cubes even faint artifacts in individual images can stack up, if present over the whole cube at a similar position, when the rotation measure synthesis technique is applied. This is also possible vice versa where strong artifacts in an individual image plane can be averaged out by rotation measure synthesis.
 Finally,  Stokes V represents the circular polarization. Astronomically circularly polarized sources are extremely rare and most often show {polarization fractions} below 1\%, so that Stokes V 
    images should nominally be regarded as empty {of 
    astrophysical signal}.

\subsection{Validation of polarization images and cubes}\label{sec:pol_qa}
{
After the overall calibration of the data,
the quality of the polarization images and cubes is mostly influenced by the instrumental leakage characteristics of the primary beam.
This means that the strongest artifacts often appear for sources far away from the beam centres where the instrumental leakage is higher. 
With this in mind we defined the following metrics, in addition to those in Section \ref{sec:cont_qa}:
\begin{itemize}
    \item $p_{in}$: The maximum value of the absolute pixel values in the inner 0.5\dg\ of the image.
    \item $FT_{max}$: The maximum of the absolute of a Fourier Transform of the image.
\end{itemize}
}

{The validation of the polarization images and cubes
provides information on their data quality;
however, the release of polarization data products is
based on the validation of the corresponding
continuum image (Section \ref{sec:cont_qa}),
as detailed in Section \ref{sec:released_proc_data}.
}
The Stokes V image and Stokes Q and U cubes {have
separate validation criteria}, due to their different nature.


\subsubsection{Validation of Stokes V images}
{
In order for a Stokes V image to pass validation
(as indicated by the data quality flag in the VO table),
the following four criteria must all be satisfied:}


\begin{enumerate}
    \item $\sigma_{in}$ and $\sigma_{out} < 60$ $\mu$Jy/beam. 
    \item $b_{min} <$ 15\arcsec. 
    \item $FT_{max} < 25$. 
    \item $p_{in} < 4$ mJy.
\end{enumerate}

{The first two criteria were chosen to ensure
minimum survey specifications were {met and mirror} the
first two criteria for continuum images. Thus, generally
a released Stokes V image will naturally pass those
criteria as the (residual) continuum image has also passed them.
Criterion 1 ensures a minimum sensitivity in the polarization images, and criterion 2 ensures the angular resolution.
}

{
The second two criteria were set to ensure a minimum of imaging
artifacts in the Stokes V images. These criteria were determined by undertaking visual examination of a large set of Stokes V images.
}
 In addition to artifacts from errors in self-calibration and direction-dependent errors
{(which are caught by the continuum image validation,
required for data release)}, leakage of strong sources near the primary beam edges also play a large role. 
{In particular, criterion 4 eliminates images
with strong instrumental leakage.
Criterion 3 is designed
to eliminate images with stripes due to unflagged RFI or bad amplitude self-calibration solutions.  }
The criteria were set so that the large majority of images which were visually assessed as good would pass validation while only a small fraction of images that were visually assessed as bad would pass.

\subsubsection{Validation of the Stokes Q and U cubes}

{In order for Stokes Q and U cubes to pass validation
(as indicated by the data quality flag in the VO table),
at least two-thirds of the image planes within the
cubes must satisfy both of the following criteria:}

\begin{enumerate}
    \item $b_{min} < 17.5$\arcsec. 
    \item $\sigma_{in}$ and $\sigma_{out} < 300$ $\mu$Jy/beam.   
\end{enumerate}

{The resolution requirement in criterion 1 is relaxed slightly compared to the continuum or Stokes V images due to the fact that the restoring beam becomes larger at lower frequencies.
The noise requirement in criterion 2 is consistent with the requirement on the Stokes V images above, assuming 24 individual images covering the same total bandwidth, giving a factor of $\sim$5 increase in noise.
}


\subsection{Overview of released polarization data quality}\label{sec:pol_valid}

{The validation of the polarization images and cubes
is not required for release but is undertaken
to provide information on their data quality.
Generally,}
the validation of the polarization images and cubes
follows that of the continuum very closely.
This is due to the use of similar metrics and
the fact that the continuum validation
already excludes the majority of polarization images and cubes
that would have large artifacts. Of the released
3374 beams, which are required to pass continuum validation (see Sections \ref{sec:released_proc_data} and \ref{sec:cont_qa}),
{only} 155 fail the Stokes V validation
and 58 fail the Stokes Q and U validation
{and are flagged as such in the VO table}. Generally,
the reason a polarized data product might fail validation
while the continuum image does not is due to a strong source at the edges of the image, where the leakage is largest.
The validation status and metrics of the polarized data products are provided in the relevant VO table (see Table \ref{tab:vo_obscore_table}).
In addition, there are 21 released beams
that have no polarized data products because a polarized
calibrator was unavailable for the observation 200309042 (S1042+5324).

The median inner and outer noise values across all Stokes V images are presented in Table \ref{tab:noise}, where the uncertainties indicate
the 16$^{th}$ and 84$^{th}$ percentiles. 
There is essentially no difference in median noise
values when considering all released {Stokes V images},
or only those that pass validation, supporting the 
release strategy {outlined in Section \ref{sec:released_proc_data} based on continuum image validation only}.

{In addition to the metrics presented here, \citetads{2022A&A...663A.103A} use polarization data from
the SVC to place a limit of 1\% on the polarization leakage
down to the 30\% response level of the primary beams.
}





\section{Line cube quality}\label{sec:hi}

The quality of line cubes is dominated by different effects
than the continuum or polarization data. While imperfect calibration or direction-dependent effects
can impact the spectral line cubes, generally artifacts in
the line cubes are dominated by two categories:
 imperfect continuum subtraction or bad frequency {ranges} (e.g., imperfect sub-bands at the system level).

It is useful to recall that four different line cubes are produced by Apercal; see 
Table \ref{tab:lineparams} for a summary of their properties. The highest frequency cube (lowest redshift),
cube3, is produced at the highest spectral resolution. Thus, it will 
have different noise properties compared to the other cubes.
The RFI environment generally worsens at lower frequencies, so cube0 will
be more strongly affected by RFI than cube1, for example.

\subsection{Validation of line cube quality}\label{sec:hi_valid}

The validation of the line cubes focused on the three lower spectral resolution cubes, as a common set of metrics could be defined for them. In practice, the quality of the high spectral resolution, high frequency cube3 follows that of the neighboring (in frequency) cube2. The following metrics were defined for each of cube0, cube1 and cube2:
\begin{itemize}
    \item $\sigma$: { The rms noise}
    \item $f_{ex}$: Fraction of the total number of pixels with an absolute value 
     $ > 6.75 \sigma $  
    \item $p_{0.8}$: Ratio between the width of the noise histogram at a level of 0.8\% of the maximum and $\sigma$  
\end{itemize}

\begin{figure*}
	\centering
	\includegraphics[width=0.9\linewidth]{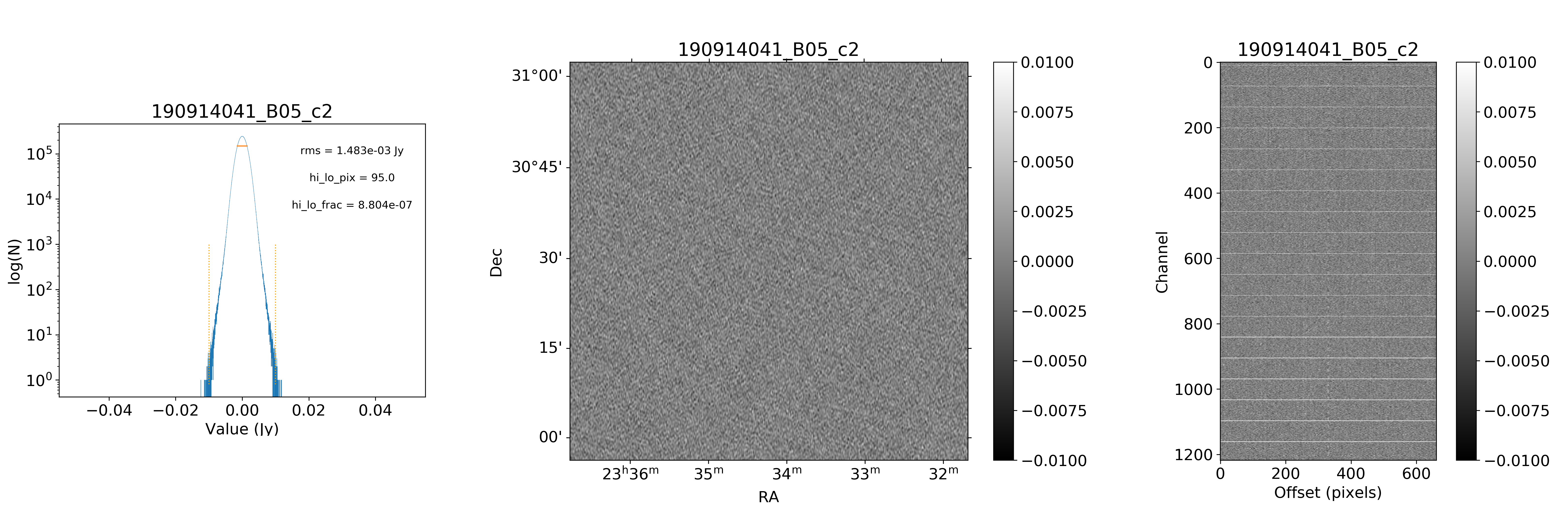}
	\includegraphics[width=0.9\linewidth]{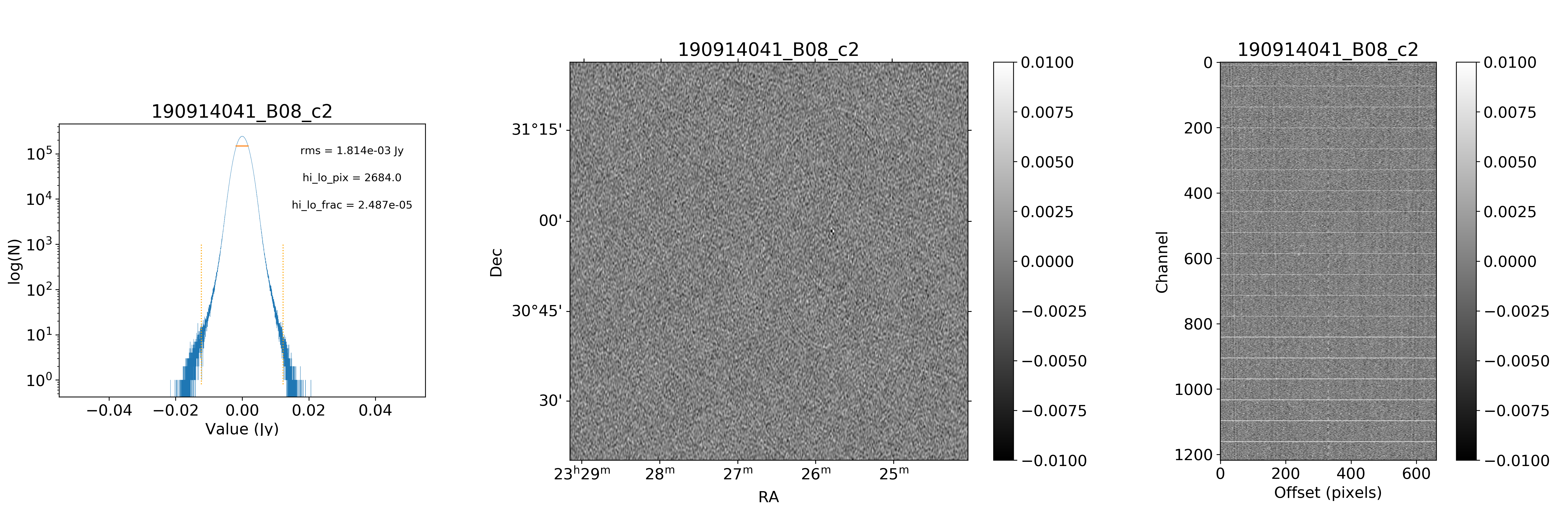}
	\includegraphics[width=0.9\linewidth]{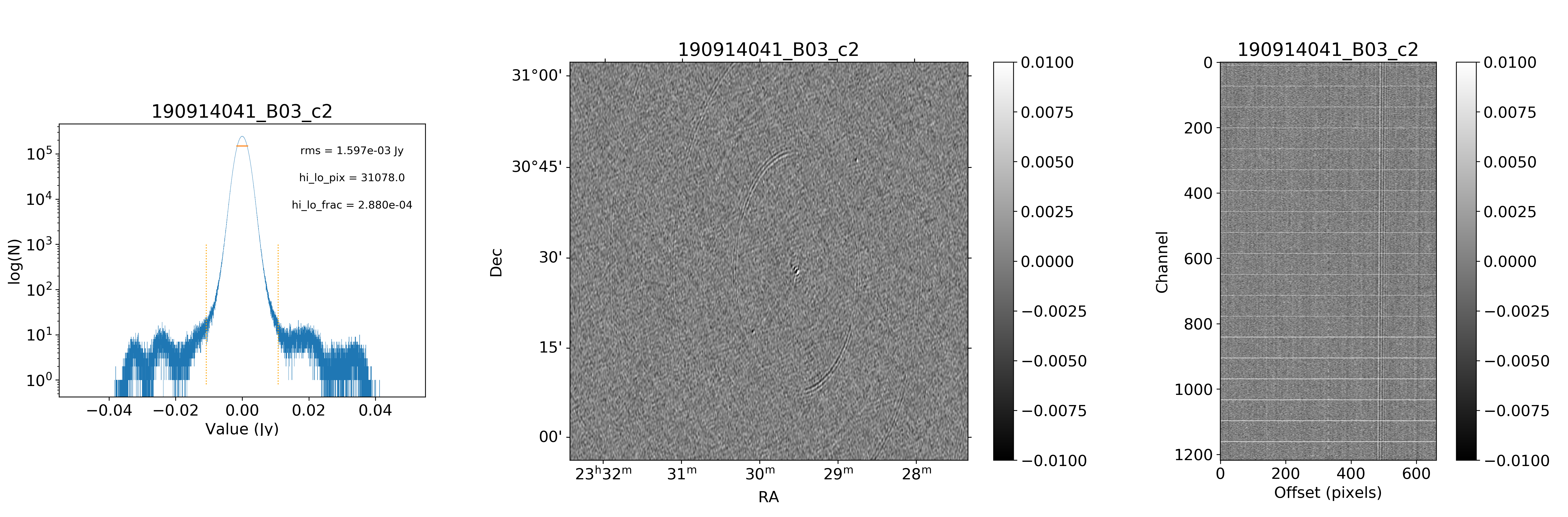}
	\caption{Examples of the three quality classes used for the HI quality
assessment. The top row shows an example of a "good" observation (ObsID
190914041, beam 5, cube 2), the middle one an "okay" observation (ObsID
190914041, beam 8, cube 2) and the bottom one a "bad" observation
(ObsID 190914041, beam 3, cube 2). 
The columns show, from left to
right, the noise histogram, an extract of the central velocity
channel, and a position-velocity diagram through the center of the
cube (the horizontal lines in the position-velocity diagrams
are the subband edges; see Section\ref{sec:aliasing} ). In the left column, the short horizontal line 
indicates {the 
rms}, $\sigma$, and
the two dotted vertical lines indicate $\pm 6.75 \sigma$.  The "good" observation in the top row shows hardly any
artifacts and a Gaussian noise histogram.  The "okay" observation in the
middle row shows a minor continuum subtraction artifact, which 
causes somewhat extended wings to the noise histogram. The "bad"
observation in the bottom row shows significant continuum subtraction
artifacts, resulting in a very non-Gaussian noise histogram.
}\label{fig:hiqa-cube2-overview}
\end{figure*}

{The last two metrics were empirically determined
to be successful at identifying deviations from Gaussianity
in the noise histograms, which correspond
to artifacts in the cubes (see Figure \ref{fig:hiqa-cube2-overview}).
}

{
Since the line cube quality can
 differ significantly from the continuum quality, upon which release of processed data is based (see Sections \ref{sec:released_proc_data} and \ref{sec:cont_qa}), this
first data release can contain cubes of poorer quality.
Given this,
rather than a strict pass/fail validation, a slightly more nuanced system of
"good", "okay" and "bad" was adopted for line cube quality:
\begin{itemize}
    \item Good: No (or very minor) artifacts 
    \item Okay: Minor artifacts present but would not significantly impact analysis
    \item Bad: Major artifacts that have to be accounted for in analysis
\end{itemize}
}

{
"Good" cubes satisfy all three of the following criteria:
}
\begin{enumerate}
    \item $\sigma$ $< 3$ mJy,
    \item $\log(f_{ex}) < -5.30$,
    \item $p_{0.8} < 0.25 f_{ex} + 5.875$.
\end{enumerate}

{
"Okay" cubes fail to fulfill all of the "good" criteria
but do meet all of the following conditions:
}
\begin{enumerate}
    \item $\sigma$ $< 3$ mJy,
    \item $-5.30 \log(f_{ex}) < -4.52$,
    \item $p_{0.8} < 0.5 f_{ex} + 7.2$.
\end{enumerate}

{
"Bad" cubes are those that do not meet the "good" or "okay" criteria. Using these conditions, we assigned all of cube0, cube1 and cube2 a ranking of "good", "okay" or "bad". Cube3 was assigned the quality of cube2 in all cases. 
}

{
In order to determine the above criteria,
a large number of cubes ($\sim$ 550 cubes from 14 observations) were visually inspected.
Figure \ref{fig:hiqa-cube2-overview} demonstrates the inspection plots,
including the noise histogram, an image of a representative channel,
and a position-velocity slice.
Generally, the first criterion ensures a minimum sensitivity
while the second two criteria identify deviations from
Gaussianity in the noise distributions of the cubes.
}

\subsection{Overview of released line cube quality}\label{sec:hi_qual}

The validation of the line cubes can diverge
quite significantly from that of the continuum images.
This is because the two main causes of line cubes failing validation, incomplete continuum subtraction and bad frequency ranges
{(whether from heavy flagging due to RFI or subbands with
poorly measured beamweights)},
can be independent of the quality of the continuum images and self-calibration\footnote{While poor continuum calibration
is likely to lead to worse continuum subtraction, good continuum calibration does not ensure good continuum subtraction. In particular, poor bandpass calibration can lead to poor continuum subtraction.}. As incomplete continuum subtraction can be addressed with a secondary subtraction in the image plane using the line cubes and bad frequency ranges are an unfortunate feature of radio observations, we do not require the line cubes to separately pass validation. Instead, the quality of the data cube ("G"ood, "O"kay, or "B"ad) is clearly indicated in the associated VO table (see Table \ref{tab:vo_obscore_table}). 
In total, 10,317 of the released cubes are good, while 1116 are okay, and 2063 are bad. Thus, over 75\% of cubes fully pass
validation while only 15\% are rated as bad.
The cubes with a quality assignment of bad may benefit from further off-line processing (i.e., continuum subtraction) before analysis.

Table \ref{tab:noise} provides a look
at the median rms values for the different spectral cubes across all the released cubes; the
uncertainties represent the $16^{th}$ and $84^{th}$ percentiles.
There is a trend that the noise values are slightly higher
for the lower frequency cubes, consistent with a worsening RFI
environment at lower frequencies,
{in addition to the fact that the system performance is generally
worse at lower frequencies}.
Limiting the assessment to only "good" cubes
decreases the median noise values slightly, as
would be expected, but the change is minor,
thus supporting the strategy of releasing all cubes where the continuum image passes validation.

{The median noise of 1.6 mJy \bm\ corresponds
to a 3-$\sigma$ column density sensitivity of
 1.8$\times 10^{20}$ atoms cm$^{-2}$ over 20 \kms; this
 is for a median declination of 38\dg\ for released
 beams, which corresponds to an angular resolution of 24\arcsec $\times$15\arcsec\ for the spectral line cubes.}



\section{Known caveats}\label{sec:caveats}

As described above, we have undertaken
validation of produced data products
in order to identify high quality data
for release.
However, there are a few global issues
that affect data quality which  we briefly
describe below.

\subsection{Ghosts}
{The Apertif data suffer
from ``ghosts''
at the center of images: bad signal with random phase
(that averages to zero).
These ghosts are most prominent in channels
16 and 48 of each subband.}
While early versions of the Apercal
pipeline flagged these channels,
this was disabled for all data
products presented here.
Before the SVC period, the finite impulse response (FIR)
filter in the channel filterbank was enabled.
{This} severely reduced the presence of ghosts
{in channels 16 and 48}, but
did not completely remove them,
{and the ghost signal was also present at a low level across
all channels}. 
Due to the reduced
presence of the ghosts {in channels 16 and 48,
plus the low level contribution from all channels}, the decision was made
to not flag channels {16 and 48 of the subbands
as the extra bandwidth outweighed the presence of the ghosts}. 
{The presence of these ghosts builds
up when averaging data over frequency, and
so they affect all Apertif data products.}
Thus, any source
identified at the exact center of a {beam}
should be treated with extreme caution.

\subsection{Aliasing}\label{sec:aliasing}

The coarse channelization of the data into subbands uses a filter
that does not have a perfectly sharp frequency response. This results in 
some overlap of response between adjacent subbands.
This effect is strongest for channels near a subband edge and also
results in a sharp drop in overall response for channels at the subband edges,
namely channels 0, 1 and 63 of every subband.
No correction is done for the aliasing, and 
the pipeline uses a brute
force approach of flagging the channels 
with suppressed signal at the subband edges.
Thus, we note that 3 out of every 64 channels are flagged.
{Since cubes0-2 have a three channel averaging
(which does not divide evenly into 64), the presence
of these flagged channels shifts around, and the 
manifestation of this flagging alternates between a single channel that is fully flagged, or two adjacent channels
that are partially flagged (1 or 2 of the
3 averaged channels)\footnote{This
accounts for the horizontal stripes in the right panels
of Figure \ref{fig:hiqa-cube2-overview}}.}
{In addition, aliased signal may occur in the presence of extremely strong
\hi\ emission,
but this will not impact the vast majority of \hi\ detections.}


\section{Scientific potential}\label{sec:science}


In this section, we briefly highlight the scientific potential
of the data contained in this first data release.

\subsection{Continuum images}
\begin{figure*}
    \centering
    \includegraphics[width=0.9\linewidth, keepaspectratio, clip=True,
    trim = 0cm 1.5cm 0cm 1.5cm]{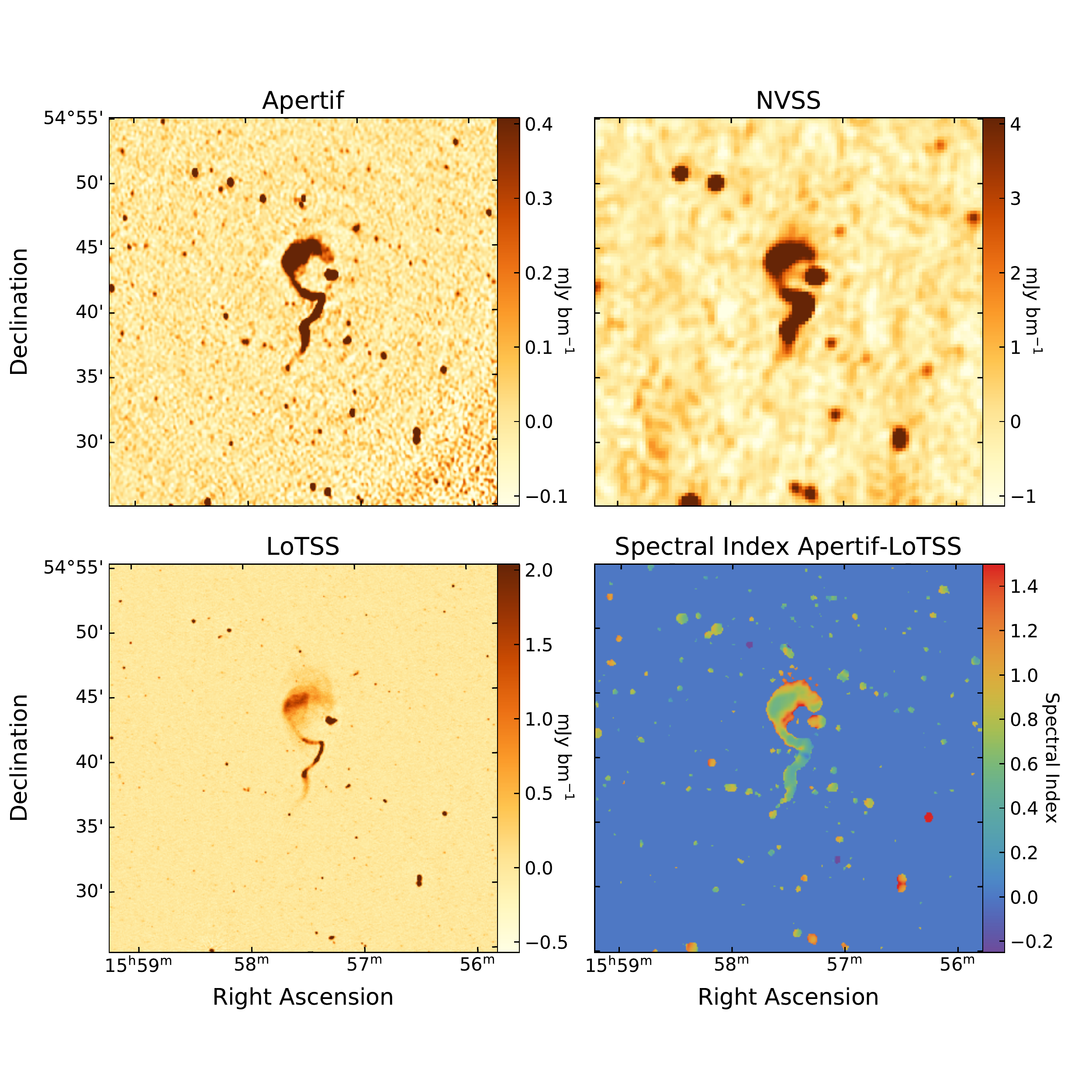}
    \caption{
    { Example Apertif continuum image with comparison to NVSS and LoTSS.
    \textit{Upper Left: } Apertif continuum image, mosaicked as described in K22.
    \textit{Upper Right:} NVSS image of the same field \citepads{1998AJ....115.1693C}.
    \textit{Lower Left:} LoTSS image of the same field \citepads{2022A&A...659A...1S}, {with the scaling set so that a point source with a spectral index of 0.7 has the same brightness as in the Apertif image}.
    \textit{Lower Right:} Spectral index image between the Apertif and LoTSS image.
    }
    }
    \label{fig:cont_example}
\end{figure*}

The Apertif continuum images have three times the angular resolution 
of NVSS and are $\sim$10 times as sensitive.
Thus, the Apertif images will both resolve more source structure and
also detect fainter emission than 
NVSS.
{ The upper two panels of Figure \ref{fig:cont_example} illustrate this, showing 
a mosaicked Apertif image (K22) and
a NVSS image over the same field of view. 
Many more point sources are seen in the Apertif data,
and the excellent surface brightness sensitivity of WSRT-Apertif
combined with angular resolution, reveal the features
of an extended, diffuse source. 
}
{ The sensitivity and resolution
of the Apertif continuum images are comparable
to those achieved by the Evolutionary Map of the Universe (EMU)
survey with ASKAP
\citepads[e.g.,][]{2022MNRAS.512.6104G};
with complementary sky coverage the two surveys augment each other
well.}

Furthermore, as described in \citetads[][see also their fig. 3]{2022A&A...658A.146V},  the depth and angular resolution of the Apertif continuum images nicely match with the images produced by the LOFAR surveys.
The synergy is particularly strong with LoTSS at 150 MHz and LoLSS at 42 ‒ 66 MHz, both aiming at covering the entire northern sky.
{ The left side of Figure \ref{fig:cont_example} demonstrates the comparison between Apertif and LoTSS data.}
The combination Apertif-LOFAR is particularly relevant for the study of the spectral properties of radio sources. 
The sensitivity of the surveys with the two instruments is well suited to trace the typical spectral index observed in extragalactic sources, around  $\alpha \sim 0.7$ or steeper\footnote{In this paper, the spectral index $\alpha$ is defined through $S \propto \nu^{-{\alpha}}$, where $S$ is flux density and $\nu$ the frequency.}. 
Specifically, given the relatively high spatial resolution, it is possible
to derive resolved spectral index images (as illustrated in the bottom right panel of Figure \ref{fig:cont_example}) for both single objects and large samples.
In particular, significant steepening of the spectrum at low frequencies reveals the presence of a particularly old electron population, possibly indicating remnant emission where the nuclear activity has stopped and the electrons have not been replenished. \citetads{morganti2021b, 2021A&A...648A...9M} presented searches for structures with such extremely steep spectral indices by combining Apertif and LOFAR 150~MHz images in the Lockman Hole region.
Finding samples of these elusive objects allows us to put {constraints} on their life-cycle \citep[see][]{jurlin2020, 2021A&A...648A...9M}.
Resolved spectral index studies are also important for diffuse emission in clusters (Orr\`u \etal in prep.,  Shulevski \etal in prep), for single objects like the radio galaxy B2~1321+31 \citepads{morganti2021b} as well as for large major mergers like  Mrk~273 \citepads{2022A&A...664A..25K}. 
These studies will be further expanded by the combination with the LBA LOFAR survey, as seen 
in the study of Mrk~273 \citepads{2022A&A...664A..25K}.

The east-west nature of WSRT means that if a source varies during
the course of an observation, it leaves a very clear pattern
in the images. \citetads{2020A&A...641L...4O} used this to identify an intra-hour 
variable during the SVC, which was then monitored as
a medium-deep field over the course of full operations.
{To date, about twenty intra-hour variables
have been identified in the Apertif survey data, along
with a few radio variable stars (Oosterloo \etal\ in prep).}

{
The image products in this data release will also complement any future independent detections of transients and variables in the area it covers.
This has recently been demonstrated for both 
fast radio bursts and for radio afterglows of gravitational-wave events.
\citet{clo+20} detected a bright fast radio burst in the Apertif time-domain survey \citep{2020A&A...635A..61O}, and then used the Apertif continuum image to set limits on 
any accompanying persistent radio source.
\citet{2021A&A...650A.131B} 
searched Apertif data for a radio counterpart to 
binary neutron-star merger gravitational-wave event GW190425. While none was found for that particular event, 
 future searches can  use DR1 as the baseline comparison for the radio-afterglow detection. 
 }

The companion paper K22 provides a continuum source catalog
based on all continuum images contained in this data release. This includes a comparison
to NVSS to identify sources that vary on long time scales, plus
a cross-match to the LoTSS DR1 to provide spectral indices for unresolved sources.

\subsection{Polarization images and cubes}

{


The Apertif polarization data, as with the continuum data,
provide an improvement of three times the angular resolution and more than ten times the sensitivity when compared to NVSS, the current benchmark
for wide surveys of the polarized sky.
The sensitivity of the wide tier results
in a polarized source density of 21 sources deg$^{-2}$ \citepads{2022A&A...663A.103A},
a factor of 20 improvement compared to NVSS \citepads{2009ApJ...702.1230T}.
This increased density of polarized sources
enables a higher angular resolution rotation measure map
to be constructed, thus providing a more detailed
look at the magnetic field of our own Milky Way galaxy.
The sensitivity of the Apertif surveys also
means that the polarized sky is being measured
down to the $\mu$Jy level.
An open question is at what level the polarized sky changes from
being dominated by AGN to star-forming galaxies;
this happens at a lower flux level than for the total
power radio continuum sky. 
\citetads{2022A&A...663A.103A}
use early Apertif data to show
that polarized source counts are still dominated by AGN,
but that the AGN are hosted in late-type, rather than early-type, galaxies.
Additionally, the high density of polarized sources
in the Apertif data provides the possibility to robustly investigate the source
counts of polarized sources and behavior of fractional polarization with total intensity (Berger \etal\ in prep).



The synergy with LOFAR, and LoTSS in particular, is also key
for polarization. 
Having information over a decade in frequency allows not only an investigation of the polarization behaviour with respect to the sources' spectral index, but also an investigation of polarization spectra. Due to Faraday Dispersion being strongly wavelength dependent, depolarisation effects are much more dominant at longer wavelengths. Combining LOFAR and Apertif data thus allows a deep insight into the nature of depolarisation mechanisms and their physical origins (Berger \etal\ in prep). 
Such an analysis also provides information on the morphology and turbulence of the magnetic fields, not only in the host objects, where the polarised emission is generated, but also along the line-of-sight to the observer. When using additional redshift information this enables an analysis of the evolution of cosmic magnetic fields \citep{2021A&A...653A.155B}.


}

\subsection{Spectral line cubes}
\begin{figure*}
    \centering
    \includegraphics[width=0.9\linewidth, clip=True, trim=0cm 2.5cm 0cm 2.5cm]{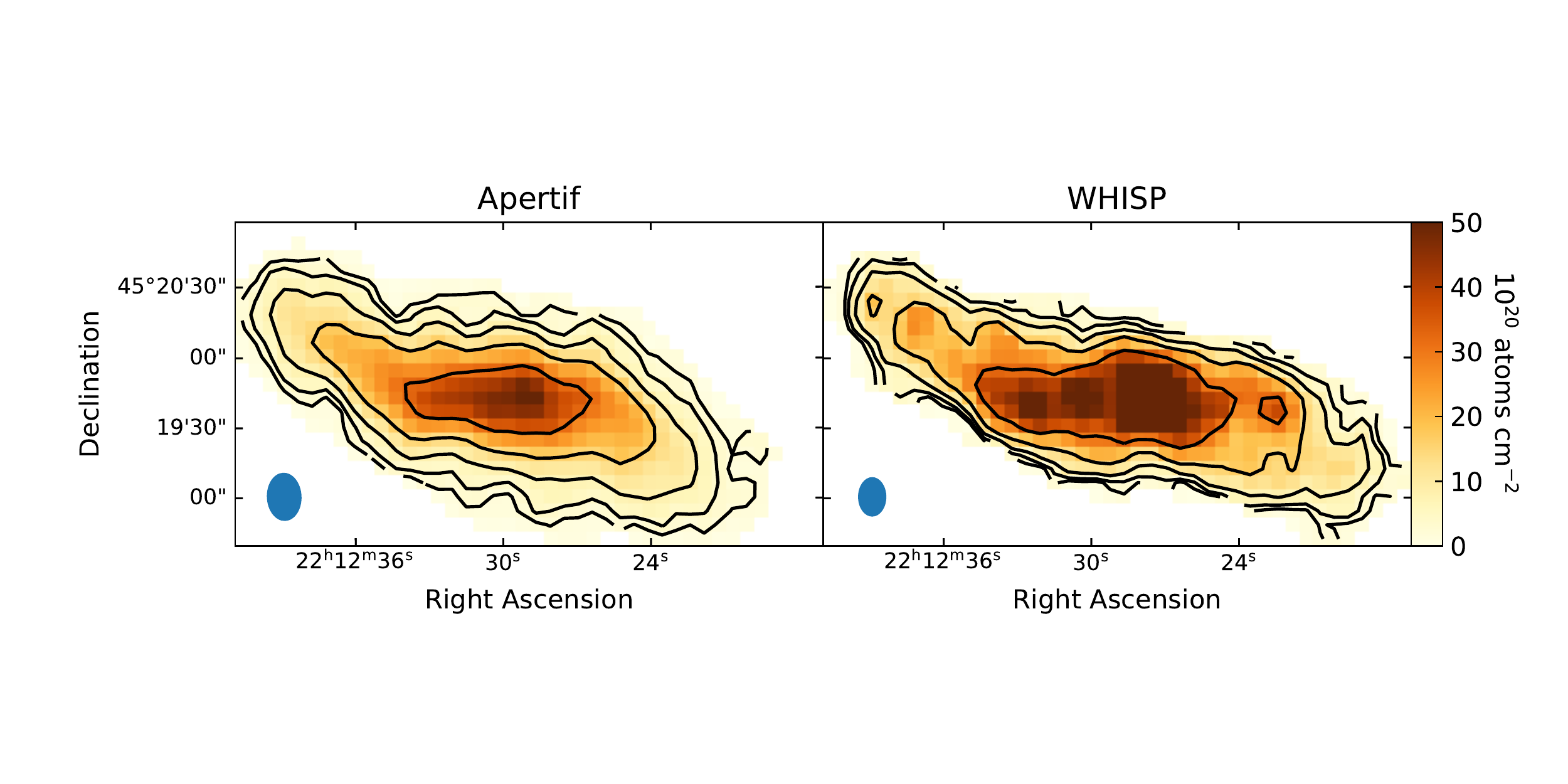}
    \caption{\hi\ content of UGC~11951 as detected in
    both Apertif (left) and WHISP (right). The color scale is the same between the two images and the contours are at [2, 4, 8, 16, 32] $\times$ $10^{20}$ atoms cm$^{-2}$.
    We note that the WHISP data have a smaller beam size as uniform weighting was used; this 
    can also account for the difference in peak column density values between the two images.
    The Apertif data is from a single beam; mosaicking would further increase the signal in that map.
    }
    \label{fig:whisp}
\end{figure*}

\begin{figure}
    \centering
    \includegraphics[width = 0.9\linewidth, keepaspectratio]{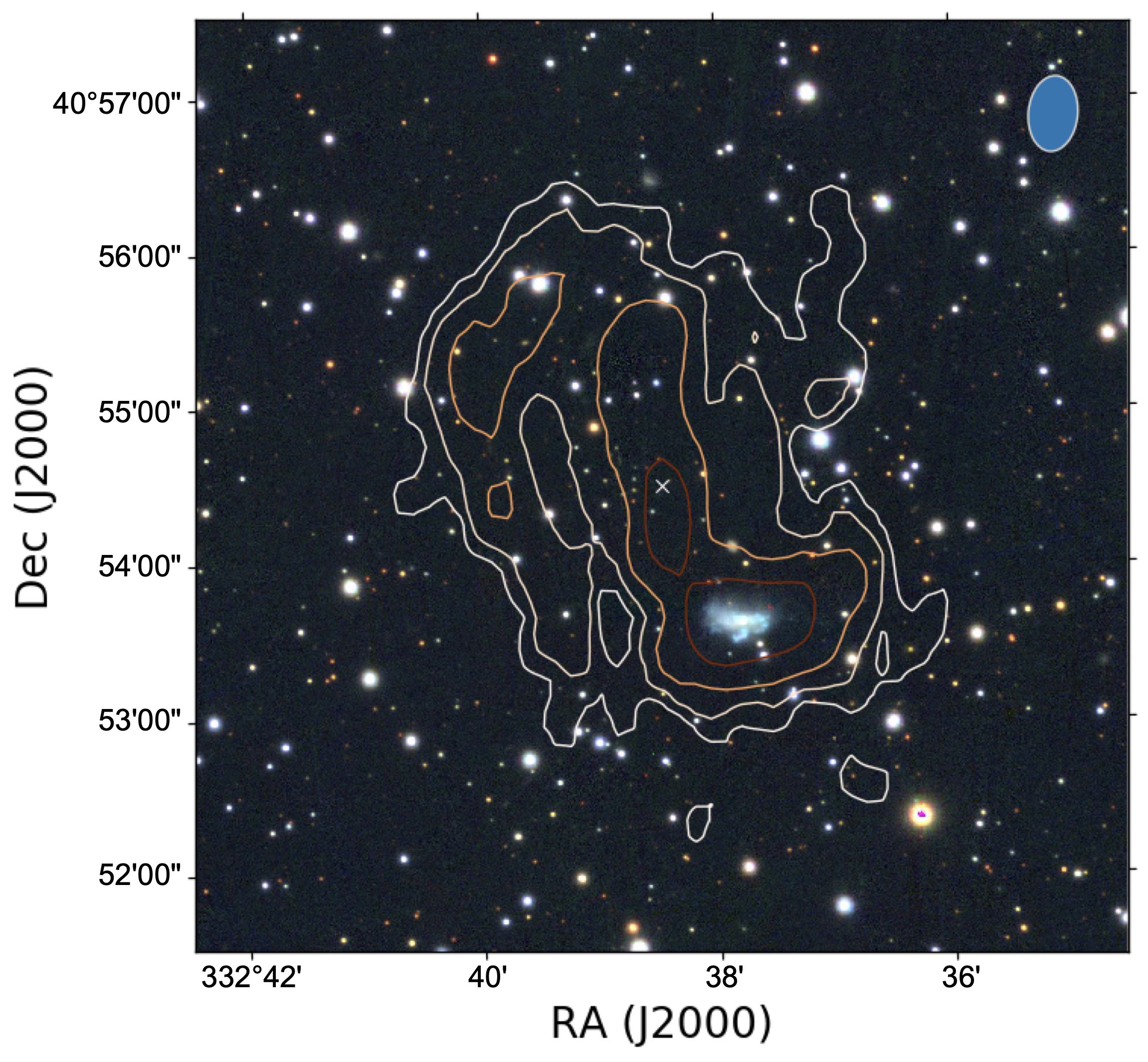}
    \caption{Example of a \hi-detected galaxy
    within the released spectral line cubes.The angular resolution provided by the Apertif surveys reveal that the \hi\ disk is hugely
    extended beyond the stellar disk. {The \hi\ contours
    are at [2, 4, 8, 16] $\times$ $10^{20}$ atoms cm$^{-2}$}}
    \label{fig:hi_example}
\end{figure}

{

The Apertif spectral line cubes offer angular resolution combined
with an untargeted detection of line emission, independent of 
the stellar content of galaxies. Compared
to single-dish \hi\ surveys, such as the 
ALFALFA \hi\ survey, 
the angular resolution improves by  $>10\times$,
allowing the distribution of the \hi\ to be studied,
providing information on disturbed morphology and environmental effects,
resolved kinematics, and more.
In terms of sensitivity and angular resolution, the wide tier is well-matched
to WHISP, a targeted survey of 375 galaxies with WSRT \citepads{2001ASPC..240..451V}. Figure \ref{fig:whisp}
shows UGC~11951 as seen in both Apertif and WHISP, where
both datasets show a warp in the \hi\ disk.
The Apertif spectral cubes offer WHISP-quality data but
for thousands of galaxies, with no selection criteria based on the stellar content.
This enables the detection of galaxies with minimal stellar content,
and the derivation of their kinematic properties to study
their behavior on galaxy scaling relations (\v{S}iljeg \etal\ in prep).
The Apertif \hi\ detections are being used to select galaxies,
based on their \hi\ morphology, for observation
with the large IFU of the new WEAVE instrument (Jin \etal\ in prep),
as part of the WEAVE-Apertif survey.
Figure \ref{fig:hi_example} demonstrates the power of the angular resolution combined with untargeted selection, showing a galaxy with a hugely extended \hi\ disk whose stellar content appears relatively undisturbed.

}

The Apertif spectral line cubes are not solely \hi\ cubes but 
OH megamasers can also be detected over their frequency coverage.
\citetads{2021A&A...647A.193H} report the serendipitous discovery of an OH megamaser
in the Apertif spectral line cubes and discuss expectations for 
the full Apertif imaging surveys, {including
how the wide bandwidth permits the simultaneous detection or placement of limits on OH megamaser satellite lines in addition to detecting the redshifted main lines}.

\section{Future prospects}\label{sec:future}

This is the first of {multiple} data releases
to come from the Apertif imaging surveys.
Improvements offered by future data releases
can be broken  into three broad categories:
\begin{enumerate}
    \item \textbf{More sky coverage} As the
    processing of the Apertif
    imaging survey data continues, there will be more processed data that can  naturally be released in the same manner as this data release.
    \item \textbf{Improved data processing}
    A major component of this release was to determine which data products were of high enough quality to  be released. Improvements to the data processing will result in more high quality data to release, improving the overall sky coverage of the public data.
    \item \textbf{Higher order data products}
    This release initially provides per-beam images and cubes. A companion
    paper K22 provides a continuum source catalog.
    The  ultimate goal is to offer more advanced science products to the community. This includes mosaicked images, but also source
    cutouts and cubelets and source catalogs for all data products 
    (polarization and spectral line). For example,
    \citetads{2022A&A...663A.103A} provide
    a polarized source catalog for the SVC observations.
\end{enumerate}

Below,  we detail a few of the key processing improvements that are 
in progress, {which we plan to incorporate in future data releases}.
\begin{itemize}

\item \textbf{Direction dependent calibration:}
In some images produced with \texttt{Apercal},
there are direction dependent errors (DDEs)
remaining around sources. We believe these
errors come from two main {sources}:
(1) pointing errors in the dishes and (2)
faulty PAF elements leading to specific compound beams on a given dish having a non-ideal {beam response} that
is different from other dishes (see \citetads{2022arXiv220509662D}
for more details). These errors cannot be 
removed in the standard self-calibration
procedure implemented in 
\texttt{Apercal}.
We took operational steps to address these errors,
updating the pointing model and targeted maintenance 
of faulty elements. However, there will always
be some bad elements, and hence compound beams with
bad shapes. Thus, we are also working on including
direction-dependent calibration in \texttt{Apercal},
where the {antenna} gains are solved in the directions
of bright sources simultaneously, pushing
the amplitude of artifacts below the thermal noise level.

\item \textbf{Two-stage continuum subtraction}
The single-stage continuum subtraction
is not  sufficient in many ($\sim$20\%) cases.
Implementing a second continuum
subtraction step will increase
the quality of the final line cubes
by minimizing the presence of
continuum artifacts.

\item \textbf{Subband-edge flagging}
The lack of
an anti-aliasing filter results in suppressed signal
on subband edges; the current approach to handling this
is to flag three of every sixty-four channels,
significantly affecting the final line cubes. 
This is likely an overly generous flagging approach 
and only the first channel of every sixty-four needs to
be flagged, while the bandpass correction can account
for the subband response in the second and last channels of each subband.

\item \textbf{Mosaicking} We are also working on a
mosaicking procedure so that in future data releases,
mosaics (both of single Apertif pointings and also across pointings) can be released in addition to individual beam images.
K22 provides an overview of a mosaicking tool {used in the creation of the continuum catalog}
that undertakes the different primary beam correction for
various Apertif beams,
including a link to source code.

\end{itemize}

\section{Summary}\label{sec:summary}

This {paper} presents the first release of data products from the Apertif imaging surveys, covering the first year of survey observations.
{The release consists of two major components: the raw observational data,
plus processed data products.
The raw observational data covers 300 MHz of bandwidth with 12.2 kHz resolution
and provide an angular resolution up to 11\arcsec/$\sin{\delta}$.}
This data release consists of 221 observations of 160 unique survey fields,
corresponding to an effective sky coverage of $\sim$1000 deg$^2$.

Processed data products available include continuum
images, Stokes V images, Stokes Q and U cubes,
and four spectral line cubes plus associated dirty beams.
{The processed data products have a bandwidth of
137.5 MHz; the lower 162.5 MHz of the band
is discarded due to strong RFI.}
We release all processed data products for beams
where the continuum image passes validation,
consisting
of 3374 observations of 2683 unique field-beam combinations.
A total of 44\% of all possible beams are released. The main reasons for continuum images to fail validation
are artifacts from poor self-calibration or direction-dependent
errors, both of which will be addressed in future versions of 
the pipeline. 

The metrics and validation status of all data products
are provided in the associated VO tables (Table \ref{tab:vo_obscore_table}). 
The continuum images have a
median inner noise of 41.4 $\mu$Jy \bm; the Stokes V images have a median inner noise of 36.9
$\mu$Jy \bm; and the spectral line cubes
have a median noise of 1.6 mJy \bm\ over 36.6 kHz.
{The median angular resolution of the continuum images and polarization data
is 11.6\arcsec/sin($\delta$).}




\begin{acknowledgements}
{We wish to thank the anonymous referee for useful comments that improved the quality {and structure} of this paper.}
      This work makes use of data from the Apertif system installed at the Westerbork Synthesis Radio Telescope owned by ASTRON. ASTRON, the Netherlands Institute for Radio Astronomy, is an institute of the Dutch Research Council (“De Nederlandse Organisatie voor Wetenschappelijk Onderzoek, NWO).
      Apertif was partly financed by the NWO Groot projects Apertif (175.010.2005.015) and Apropos (175.010.2009.012).
      This work was partly supported by funding from the European Research Council under
    the European Union’s Seventh Framework Programme (FP/2007-2013), through
    ERC Grant Agreement No. 291531 (‘HIStoryNU’, PI: JMvdH) and ERC Advanced
    Grant RADIOLIFE-320745 (PI: RM), in addition to funding from NWO via
    grant TOP1EW.14.105 (PI: TAO).
    EAKA is supported by the WISE research programme, which is financed by NWO. 
    BA acknowledges funding from the German Science Foundation DFG, within the Collaborative Research Center SFB1491 ``Cosmic Interacting Matters - From Source to Signal''.
    KMH acknowledges financial support from the State Agency for Research of the Spanish Ministry of Science, Innovation and Universities through the "Center of Excellence Severo Ochoa" awarded to the Instituto de Astrofísica de Andalucía (SEV-2017-0709), from the coordination of the participation in SKA-SPAIN, funded by the Ministry of Science and Innovation (MCIN). OMB and JvL acknowledge funding from NWO under the Vici research program 'ARGO' with project number 639.043.815. YM, LCO, RS and JvL acknowledge funding from the European Research Council under the European Union's Seventh Framework Programme (FP/2007-2013)/ERC Grant Agreement No. 617199 ('ALERT').
IPM acknowledges funding from the Netherlands Research School for Astronomy (grant no. NOVA5-NW3-10.3.5.14). AAP acknowledges support of the STFC consolidated grant ST/S000488/1. 
DV acknowledges support from the Netherlands eScience Center (NLeSC) under grant ASDI.15.406.

This research has made use of NASA’s Astrophysics Data System Bibliographic Services and Astropy,\footnote{http://www.astropy.org} a community-developed core Python package for Astronomy \citep{astropy:2013, astropy:2018}
\end{acknowledgements}



\bibliographystyle{aa}
\bibliography{refs}


\appendix

\section{Additional available data}\label{app:otherdata}

In addition to the release of survey data,
imaging data from before the start of survey operations
are also available. 
These data are not hosted via VO tables but can be accessed via requests to the ASTRON
helpdesk\footnote{\url{https://support.astron.nl/jira/servicedesk}}.

\subsection{Data from the science verification campaign}\label{app:svc}
Before the start of survey operations,
the SVC
was undertaken to verify the scientific
performance of the Apertif system.
The goal of this period was to take data in {as close to final
survey mode as possible} to verify the science performance of Apertif before
the start of surveys.
The good quality data from this period,
including processed imaging
data products, were publicly released
at the end of 2019. In this Appendix 
we briefly describe the imaging observations
and data quality that
are part of this earlier release.

\subsubsection{Observations and processing}
The target fields and associated
calibrators for the SVC period
are listed in Table \ref{tab:svc_fields}.
During the SVC period, 
the observing frequency range
was 1250-1550 MHz. 
The data was processed
with a 150 MHz version of the pipeline
specific to the SVC period\footnote{
The SVC specific version of the pipeline is available
at https://github.com/apertif/apercal/releases/tag/v2.4}. The frequency
range of the final data products is close,
but not quite
identical to that used for the survey data products (1291.8–1441.8 MHz for the SVC data).
However, the line cubes
are produced over the same frequency range
(given in Table \ref{tab:lineparams}).
One key processing difference compared to the survey processing is that the SVC data required an extra offline correction for the delay tracking; this correction
is done online on the datawriter during survey observations (see Section 8.3 of \citetads{2022A&A...658A.146V}).

\subsubsection{SVC data quality}\label{sec:svc_qual}
The processed SVC data was released
without validation as a demonstrator of
data quality, both the intrinsic
quality and pipeline processing.
For comparison to the survey data release,
Table \ref{tab:svc_qa}
provides a high-level overview
of the number of beams that would
have passed continuum validation for each field,
if the validation process as for the survey
data release had been used.

\subsection{Early science observations}\label{app:earlyscience}

There was a $\sim$2.5 month period
between the SVC and start of survey operations.
This period was focused on development and finalizing
tools for  operations but did offer the  possibility
for high quality early science observations.
These observations are listed in Table \ref{tab:early_science_fields}. 
These data were  observed at different frequency settings and only the last   observations
have the online correction for delay tracking.
We  are releasing these observations but not
associated processed data; none of these observations
have been fully processed  by the Apercal pipeline.
We do note that one field (ObsID 190428055) was calibrated with 
Apercal and manually imaged and published in \citetads{2021A&A...648A...9M}.

\begin{table}[]
    \centering
    \caption{Summary of continuum validation for SVC beams}
    \label{tab:svc_qa}
    \begin{tabular}{l|l|l}
    \hline \hline
         Field & Pass & Fail \\
        \hline
         S2248+33\tablefootmark{a} & 26 & 3\\
         M1403+53\tablefootmark{b} & 4 & 34 \\
         M0155+33\tablefootmark{b} & 6 & 32 \\
         S2246+38\tablefootmark{b} & 15 & 23 \\
         S1415+36\tablefootmark{b} & 24 & 14\\
    \hline
    \end{tabular}
    \tablefoot{
    \tablefootmark{a}{Twenty-nine beams processed}
    \tablefootmark{b}{Thirty-eight beams processed.}
    }
\end{table}

\begin{table*}[]
    \centering
    \caption{Overview of the imaging SVC fields}
    \label{tab:svc_fields}
    \begin{tabular}{ll|lll|lll}
    \hline \hline
    \multicolumn{2}{c}{Field}  & \multicolumn{3}{c}{Flux calibrator} &
    \multicolumn{3}{c}{Polarization calibrator} \\
    Name & Task ID & Name & Task IDs & t$_{obs}$ & Name & Task IDs & t$_{obs}$ \\
         &         &      &          & min       &      &          & min \\
     \hline
    S2248+33\tablefootmark{a} & 190409015 & 3C196  & 190408125-150 & 5 &
    3C138 & 190409016-055 & 4 \\
     &                &     & 190409001-014\tablefootmark{a} &  & & & \\ 
    M1403+53 & 190409056 &3C196 & 190410002-041 & 5 &3C138 & 190409016-055 & 4 \\
    M0155+33 & 190410001 &3C196 & 190410002-041 & 5 &3C138 & 190409016-055 & 4  \\
    S2246+38 & 190411001 &3C196 & 190410002-041 & 5 &3C138 & 190411002-041 & 4  \\
    S1415+36 & 190411042 &3C196\tablefootmark{b} & 190410002-041\tablefootmark{b} & 5 &3C138 & 190411002-041 & 4 \\
    \hline
    \end{tabular}
    \tablefoot{
    \tablefootmark{a}{Beams 31-39 failed (190409006-14 not on source) }
    \tablefootmark{b}{Non-bracketing flux calibrator used due to failure of observing session}
    }
\end{table*}

\begin{table*}[]
    \centering
    \caption{Overview of the imaging early science fields}
    \label{tab:early_science_fields}
    \begin{tabular}{ll|ll|ll|llp{2.2cm}}
    \hline \hline
    \multicolumn{2}{c}{Field}  & \multicolumn{2}{c}{Flux calibrator} &
    \multicolumn{2}{c}{Polarization calibrator} & & \\
    \hline
    Name & ObsID &   Name & ObsIDs & Name & ObsIDs & Freq   &DT\tablefootmark{a}& Notes\\
         &         &     &      &  & & MHz    & &  \\
     \hline
    1530+29 & 190419137 &  3C147  & 190419097$-$136  &
    3C138 & 190420001$-$40 &1250$-$1550 & N  & RTC\&RTD bad  \\
    1530+29 & 190424047  & 3C147 &  190424007$-$46 & -- &-- &  1250$-$1550 & N  & RTC\&RTD bad\\
    LH\_WSRT & 190428055 & 3C147 & 190428016$-$54 & 3C286  & 190429001$-$40 & 1250$-$1550 & N  &\\
    LH\_WSRT & 190505048   & 3C147  &  190505008$-$47  & 3C286 & 190506001$-$40 & 1220$-$1520  &  N & \\
    M31  & 190511013  &  3C147 & 190512001$-$40 & 3C286 & 190510107$-$190511012 & 1220$-$1520 & N  & \\
    M1403+53 & 190511014 & 3C147& 190512001$-$40 & 3C286 & 190510107$-$190511012 & 1220$-$1520  & N  &  \\
    LH\_GRG & 190601059  & 3C147  & 190601019$-$58 & -- &  -- & 1220$-$1520 & N & \\
    LH\_GRG & 190602049  & 3C147 & 190602009$-$48 & 3C286 & 190602050$-$190603031 & 1130$-$1430 & N & \\
    M1403+53 & 190608061  & 3C147 & 190609001$-$40 &  3C138 & 190608021$-$60 & 1220$-$1520 & N & no  RTD; delay center offset = [0.33,0] \\
    M81 &  190609041 & 3C48 & 190610001$-$40 & 3C138 & 190610041$-$80 & 1220$-$1520 & N & no RTD \\
    M81\_offset2 & 190610081 & 3C48 & 190610001$-$40 & 3C138 & 190610041$-$80  & 1220$-$1520 & N & no RTD \\
    S2258+29 & 190629059   & 3C147 & 190629018$-$57 & 3C138  & 190630001-40 & $1130-1430$ &  Y & \\
    S1349+26  & 190630041  &  3C147 & 190701001$-$41 & 3C138 & 190630001$-$40 &  1130$-$1430 & Y &\\
    M0142+31 & 190701001 & 3C147 &  190701002$-$41 & 3C138 & 190630001-40  & 1130$-$1430 & Y &\\

    \hline
    \end{tabular}
    \tablefoot{
    \tablefootmark{a}{Whether online delay tracking (DT) was
    active or not.}
    }
\end{table*}

\end{document}